\documentclass[5p,twocolumn,number]{elsarticle}
\usepackage{datetime}
\usepackage{amssymb, amsmath}
\usepackage{color}
\usepackage{xspace}
\usepackage{booktabs}
\usepackage{gensymb}
\usepackage{placeins}
\usepackage{eso-pic}
\usepackage{setspace}
\usepackage[rightcaption]{sidecap}
\sidecaptionvpos{figure}{c}
\usepackage[]{hyperref}
\definecolor{DarkBlue}{rgb}{0.00,0.00,0.75}
\hypersetup{
  linkcolor   = DarkBlue,
  anchorcolor = DarkBlue,
  citecolor   = DarkBlue,
  filecolor   = DarkBlue,
  urlcolor    = DarkBlue,
  colorlinks  = true,
  pdftitle    = {An implicit wetting and drying approach for non-hydrostatic baroclinic flows in high aspect ratio domains},
  pdfauthor   = {Dr Adam S. Candy},
}
\usepackage{cleveref}
\biboptions{sort&compress}
\renewcommand{\today}{Monday 23\textsuperscript{rd} January, 2017}

\providecommand{\abs}[1]{\left|#1\right|}
\providecommand{\norm}[1]{\left\lVert#1\right\rVert}
\providecommand{\eqref}[1]{(\ref{#1})}
\newcommand{\bcdot}{\boldsymbol{\cdot}}
\newcommand{\bnabla}{\boldsymbol{\nabla}}
\newcommand{\ud}{\,\mathrm{d}}

\newcommand{\popo}{\ensuremath{P_{1}-P_{1}} }
\newcommand{\podgpt}{\ensuremath{P_{1}^\mathrm{DG}-P_{2}}\xspace}
\newcommand{\podg}{\ensuremath{P_{1}^\mathrm{DG}}\xspace}
\newcommand{\po}{\ensuremath{P_{1}}\xspace}
\newcommand{\fp}{\ensuremath{p(\overline{p},\eta)}\xspace}
\newcommand{\twod}{2D\xspace}
\newcommand{\threed}{3D\xspace}
\newcommand{\ts}{\ensuremath{\Delta t}\xspace}

\crefname{equation}{}{}
\crefname{figure}{figure}{figures}
\usepackage{titlesec}
\titlespacing\section{0pt}{5pt plus 4pt minus 2pt}{2pt plus 2pt minus 2pt}
\titlespacing\subsection{0pt}{5pt plus 4pt minus 2pt}{2pt plus 2pt minus 2pt}
\titlespacing\subsubsection{0pt}{5pt plus 4pt minus 2pt}{2pt plus 2pt minus 2pt}
\journal{Advances in Water Resources}
\begin{document}
\setlength{\abovedisplayskip}{5pt plus 2pt minus 2pt}
\setlength{\belowdisplayskip}{5pt plus 2pt minus 2pt}
\begin{frontmatter}
\title{An implicit wetting and drying approach for non-hydrostatic \\ baroclinic flows in high aspect ratio domains}
\author[ese,tud]{Adam~S.~Candy\corref{cor}}
\ead{a.s.candy@tudelft.nl}
\cortext[cor]{Corresponding author address:
Adam S. Candy, now at:
Environmental Fluid Mechanics Section, Faculty of Civil Engineering and Geosciences, Delft University of Technology, The Netherlands.}
\address[ese]{Department of Earth Science and Engineering, Imperial College London, UK}
\address[tud]{Environmental Fluid Mechanics Section, Faculty of Civil Engineering and Geosciences, Delft University of Technology, The
Netherlands}

\begin{abstract}
A new approach to modelling free surface flows is developed that enables, for the first time,
\threed consistent non-hydrostatic baroclinic physics that wets and dries in the large aspect ratio spatial domains that characterise geophysical systems.
This is key in the integration of physical models to permit seamless simulation in a single consistent arbitrarily unstructured multiscale and multi-physics dynamical model.
A high order continuum representation is achieved through a general Galerkin finite element formulation that guarantees local and global mass conservation,
and consistent tracer advection.
A flexible spatial discretisation permits conforming domain bounds and a variable spatial resolution, whilst
atypical use of fully implicit time integration ensures computational efficiency.
Notably this brings the natural inclusion of non-hydrostatic baroclinic physics and a consideration of vertical inertia to flood modelling
in the full \threed domain.
This has application in improving modelling of inundation processes in geophysical domains, where dynamics proceeds over a large range of horizontal extents relative to vertical resolution,
such as in the evolution of a tsunami,
or in urban environments containing complex geometric structures at a range of scales.
\end{abstract}
\begin{keyword}
Wetting and drying \sep
Non-hydrostatic \sep
Baroclinic \sep
High aspect ratio domains \sep
Multi-scale simulation\sep
Vertical inertia\sep
Finite element method
\end{keyword}
\end{frontmatter}

\section{Introduction}
\label{introduction}
Flooding has huge impacts on the economy of a region and the livelihood of its people.
Significant progress has been made to model and predict the impact of these inundation events,
capturing the character of their source and resultant behaviour.
Many challenges still exist and in particular in concurrently simulating
the physical processes involved from the large planet-scale forcings down to the small human scales of an urban environment.
This is highlighted in the review \cite{medeiros13} as one of the key limitations of existing wetting and drying (WD) models.
In an urban flooding scenario for example,
modelled water column depth could be down to $1\textrm{cm}$ over a horizontal range of tens or hundreds of kilometres,
leading to a very high aspect ratio of ${\sim\!{10}^{-7}}$.

Inundation flow models typically use simplified formulations of the Navier-Stokes equations, commonly the Saint-Venant shallow water equations (SWEs).
These simplifications make assumptions, such as a hydrostatic pressure and well-mixed water column, which are not necessarily valid in the whole range of scales relevant to the inundation.
Non-hydrostatic processes become important, for example,
in the dispersive effects of short waves where the ratio of vertical and horizontal scales of motion are not sufficiently small.
The study of \cite{onishi13}, considering the 2011 T\={o}hoku tsunami in Japan, found it was critical to include non-hydrostatic effects to correctly model processes on the small scale, a point further highlighted by \cite{cui14}.

Multi-physics over a broad range of scales is typically approached using multiple model runs at a hierarchy of scales such that domains are nested,
with varying complexity and included physics.
As an alternative, efforts to integrate the physics and scales of separate models into single Earth system models is growing, where it is important individual components function in a general context, and are not too restrictive in discretisation choice.
Although this can be achieved weakly with offline communication between models, the `holy grail' is a flexible single model capable of simulating a range of physics and scales, with inherit consistency and conservation.

\begin{figure*}
\centering
\includegraphics[width=17cm]{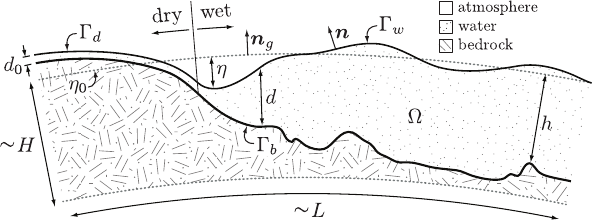}
\vspace{-0.3cm}
\caption{
A schematic of an example high aspect ratio geophysical inundation domain considered here,
with a `horizontal' length scale $L$ spanning its extent on a geoid surface, and the `vertical' length scale $H$.
In reality, these length scales differ by many orders of magnitude.
}
\label{fig:domain}
\end{figure*}

This work pushes the boundaries in two key regards:
firstly, adding a novel approach to WD in the `thin-film' family solving a full \threed pressure rather than the usual SWE approximation
in very challenging acutely large aspect ratio domains typical of geophysical systems {\textemdash} a first for WD.
Secondly, this
brings the modelling of WD processes together with non-hydrostatic baroclinic flow dynamics in a single simultaneous and seamless system model.
This is critical for tightly coupled processes, for example in tracking grounding line movement under an ice shelf ocean cavity, that is strongly influenced by non-hydrostatic and baroclinic ocean flows.

Accurately tracking an inundation interface is technically challenging.
Of the Eulerian type WD approaches \citep[reviewed in][]{medeiros13},
where the underlying spatial discretisation is predominantly independent of space and time
such that matrix operators can be cached and there is no need for complex contour tracking,
four types exist:
element removal, limiting the computational domain to the wet region
(see \cite{casulli98}, UnTRIM \cite{casulli00}, \cite{defina00,dalpaos07} and the WASH123D code \cite{lin04});
thin film approaches
(see \cite{bates93}, the FVCOM model \cite{chen03}, POM model \cite{oey05} and also
\cite{begnudelli06});
depth extrapolation from wet to dry cells
\cite{bradford02,lynett02};
and negative depth
\cite{heniche00, jiang05} applied in ROMS \cite{warner13},
including the use of porous media below the sea bed
\cite{vant05,ip98} and bathymetry  movement
\cite{karna11}.

Underlying model discretisations largely steer this choice,
with the first by far the most common for explicit time stepping models,
applied in QUODDY, ADCIRC, MIKE21, Delft3D and in one of the first Eulerian methods \cite{leendertse70}, subsequently reviewed in \cite{balzano98}.
Whilst robust,
stability constraints restrict movement of the interface to one cell per time step (\ts),
since the Courant number in drying regions must be maintained less than one \cite{stelling03a} to ensure a non-negative bound on water depth,
a strict limitation on \ts
\cite{walters05}.
Depth extrapolation also suffers this restriction with elements switching
states \cite{medeiros13}, whereas thin film and negative depth options can be time-integrated implicitly.

For spatial discretisations,
WD procedures were first applied to structured meshes
\cite{casulli98,stelling03a},
with updates to include non-hydrostatic corrections
\cite{stelling03b},
baroclinic solvers \cite{warner13}
and recently subgrid information
\cite{defina00,dalpaos07,casulli10,volp16}
to include higher resolution bathymetry and flux calculations.

Current approaches to unstructured mesh geophysical fluid modelling are considered in detail in \cite{danilov13},
with their potential importance best highlighted in \cite{danilov13c}.
Indeed, this review states that whilst unstructured mesh models may not replace structured modelling approaches completely, there are cases where this type of approach could be optimal.
In particular, allowing a flexible approach to the vertical discretisation could improve accuracy and model efficiency in domains where there are sharp changes in bathymetry relative to horizontal spatial resolution,
strong non-hydrostatic gradients in pressure,
strong vertical inertial flows,
or when it would be more optimal to reduce or increase the number of layers in shallow and deep regions, respectively.
Moreover, these could be critical in the fringes of the ocean boundary, along geometrically complex coastlines and in interactions with other types of physical systems, such as an urban environment or the complex shallowing in ice shelf ocean cavities.
Within this discretisation type, WD models can more accurately model a wider range of scales in larger domains.

One of the early finite volume (FV) approaches UnTRIM \cite{casulli00} permits unstructured meshes
with the constraint that, like structured models, the domain elements are
orthogonal
where circumcentres
are inside their respective elements.
Its non-hydrostatic advance
\cite{casulli02} is applied in
the SUNTANS model,
with the same orthogonality restriction.
It contains a WD algorithm \cite{wang09}
stabilised with a technique from \cite{ip98} that applies an increased drag
to satisfy an additional constraint on volume flues in dry regions.
Similar approaches are also made in
\citep[][FVCOM, MIKE21]{cui10}
with non-hydrostatic corrections added
\citep[e.g.][]{cui12}.
FV is low order only and
models generally explicit.

Unstructured finite element (FE) methods offer high order continuum approximations
which are more accurate and naturally include less diffusive and dispersive advection schemes.
WD has been built into \twod barotropic flow FE models such as
QUODDY with dry element removal in \cite{greenberg05};
ADCIRC, a SWE method for explicit hydrostatic modelling of storm surges with dry removal \cite{dietrich06};
TELEMAC, initially using element removal \cite{bates99} and now negative depth;
and
SLIM \cite{karna11} with a repositioned sea bed SWE method
and
adoption of implicit \ts advance.

WD is combined with solvers capable of modelling baroclinic processes in \cite{funke11,warner13}, with the former using thin film high order FE and the latter explicit finite difference with negative depth WD and mode splitting.
The former performs well in
relatively modest aspect ratio domains
but performance is strictly limited by the following:
use of direct solvers (LU decomposition),
restrictions on dry element aspect ratios
and erroneous unphysical flows that develop in dry regions.

Here a general approach for WD with FEs is considered in full \threed,
building on established methods for modelling fluid flow on fully \threed unstructured meshes~\cite{piggott08} which vary in resolution
and support a multiscale of physical processes,
including non-hydrostatic and baroclinic dynamics
in the large aspect ratio domains found in geophysical domains.
Under the constraints of a global number of degrees of freedom, this allows the focus of computational resources on small scale regions and areas of interest, whilst capturing the large scale flows elsewhere in the domain.
An additional advantage is that there is neither a constraint on the internal mesh structure, nor is it fixed in time.
It is not constrained to layers, and can be completely (or partially in select regions) fully anisotropically unstructured.

To allow efficient time integration over a range of element sizes, an implicit treatment necessitates a continuum approach to interface tracking.
A thin film is applied, which as \cite{medeiros13} notes,
generally satisfies mass and momentum conservation
without significant special treatment, and produces a realistic and smooth wetting front.
WD is included in a natural manner, through additional terms in the momentum equation and modified boundary conditions.
Indeed, the numerical treatment is careful to ensure the solution remains in the Sobolev solution space of the original physically-based weakly formulated Galerkin problem.
Prognostic variables, including tracers, are self-consistent through the FE formulation
and notably, through use of a combined pressure variable, consistency with the free surface is naturally inherent.

In the following \cref{sec:equations,sec:discetisation,sec:conditioning},
the new consistent approach for
WD in large aspect ratio geophysical domains
is developed,
with
details of mesh movement in
\cref{sec:meshmovement} and
additional strategies noted in
\cref{sec:additional}.
This is validated in \cref{sec:application} and evaluated in \cref{sec:conclusions}.

\section{Governing continuum equations}
\label{sec:equations}
\subsection{\threed Boussinesq with piezometric pressure}
The non-hydrostatic Boussinesq equations for a rotating stratified fluid,
are solved in a
time-dependent domain $\Omega \subset \mathbb{R}^3$
(see \cref{fig:domain}),
bounded by the surface $\Gamma$.
This is split into the free surface boundary $\Gamma_\eta$, and the remaining bound $\Gamma_b = \Gamma \setminus \Gamma_\eta$.
These are defined for the
prognostic variables of velocity
$\boldsymbol{u}\!: \Omega \times [0,T)\mapsto \mathbb{R}^{3}$,
and
pressure
$\overline{p}\!: \Omega \times [0,T) \mapsto \mathbb{R}$,
over the time interval [0, T),
such that
\begin{align}
\rho_0 \left(\dfrac {\partial \boldsymbol{u}} {\partial t} + \boldsymbol{u} \bcdot \bnabla \boldsymbol{u} \right)
- \bnabla \bcdot \mu \bnabla \boldsymbol{u} + \bnabla p
&= - g \rho' \boldsymbol{n}_{g},
\label{momentumequation_nohydrostatic}
\\
\bnabla \bcdot \boldsymbol{u}
&=0,
\label{incompressibility2}
\end{align}
where
$\mu$ is the tensorial dynamic viscosity,
$-\boldsymbol{n}_g$
and $g$ the gravitational acceleration direction and magnitude respectively, and
$\rho\!:\Omega \times [0,T)\mapsto \mathbb{R}$
the density. The latter is split into a background $\rho_0$,
and perturbation density $\rho'$,
such that $\rho = \rho_0 + \rho'$.
Since the hydrostatic component of pressure of the equilibrium state does not have an important contribution dynamically, it is subtracted from the momentum equation and the full pressure $\overline{p}$, is replaced by
a \emph{piezometric} pressure,
commonly applied in coastal engineering applications
\citep[e.g.][]{labeur05},
defined as
\begin{equation}
p\!:= \overline{p} + \rho_0 g \boldsymbol{n}_g \bcdot \boldsymbol{r} + p_a,
\label{piezometric}
\end{equation}
for a position vector $\boldsymbol{r}$, relative to a position where hydrostatic pressure is zero.
Atmospheric pressure at the interface is denoted $p_a$.

Redefining the prognostic pressure with this particular choice of piezometric pressure
forms a combined
free surface~--~pressure
prognostic \fp
eliminating the need to solve a separate, commonly used, wave equation for the free surface,
denoted by the injective function $\eta\!: \Omega \times [0,T) \mapsto \Gamma_\eta$.
The prognostic pressure $p$ now contains non-hydrostatic components
and the hydrostatic pressure due to perturbations in the free surface elevation.
This remaining hydrostatic pressure $\rho_0 g\eta$, is the boundary condition for $p$ at $\Gamma_\eta$,
and through \cref{piezometric},
we find
\begin{equation}
p \, \big|_{\Gamma_\eta}
= \rho_0 g\eta.
\label{pressurebc}
\end{equation}

\subsection{Boundary conditions}
With the inclusion of the free surface height in the prognostic pressure, the kinematic
free surface
boundary condition
of \cref{sec:kinematic}
is expressed
\begin{equation}
\boldsymbol{n} \bcdot \boldsymbol{n}_g
\left. \frac{\partial p}{\partial t}
\right|_{\Gamma_\eta}
=
\rho_0 g \,
\boldsymbol{n} \bcdot \boldsymbol{u},
\quad \textrm{on} \; \Gamma_\eta.
\label{FSeq1-1b}
\end{equation}
This is the boundary condition for $\eta$
and now a required constraint for the combined
\fp prognostic variable.
This is joined by the
$\boldsymbol{u}$ and $p$
boundary constraints
\begin{align}
& \boldsymbol{u} \bcdot \boldsymbol{n} = 0, \; \forall \boldsymbol{x} \in \Gamma_b, \; \textrm{and}
\nonumber
\\
& p = p_a, \; \forall \boldsymbol{x} \in \Gamma_\eta.
\label{boundaryconditions}
\end{align}
More general conditions, for open ocean boundaries or flux inputs, can be applied without fundamental changes to the approach.

\subsection{Coordinate system and frame of reference}
Note additionally that the direction of gravity, describing the normal $\boldsymbol{n}_g$,
is not restricted to a Cartesian $z$-component, such that the development is relatively independent of the coordinate reference frame.
It is free to vary arbitrarily within $\mathbb{R}^3$,
aligned with the local direction of gravitational acceleration,
and it is possible for example, to apply this method to the
spheroid shell of the Earth in a Cartesian coordinate reference frame.

\section{Spatial and temporal discretisation}
\label{sec:discetisation}
\label{sec:discretefreesurface}
The non-linear system of equations~\cref{momentumequation_nohydrostatic,incompressibility2},
combined with
boundary conditions
\cref{FSeq1-1b,boundaryconditions},
are solved for \fp, and velocity $\boldsymbol{u}$,
using a
Chorin
projection method
\cite{chorin67}
to enforce incompressibility.
This is a modified predictor~--~corrector scheme based on
\cite{gresho84} in which a predictor $\boldsymbol{u}_*^{n+1}$
is obtained from momentum conservation that is not divergence free,
such that a correction
$\boldsymbol{u}^{n+1} = \boldsymbol{u}^{n+1}_* - \bnabla \phi$
is then calculated subject to the divergence-free constraint
$\bnabla \bcdot \boldsymbol{u}^{n+1} = 0$.
For each time step, this proceeds for a number of Picard iterations until sufficiently converged.

\subsection{Temporal discretisation}
\label{sec:temporal}
Discretisation in time is achieved by the
$\theta$-method~\cite{iserles12} in all cases,
for a time step $\Delta t$, such that the explicit forward Euler, Crank-Nicolson and backward Euler time-stepping schemes can be obtained with choices of $\theta = 0$, $\frac{1}{2}$ and $1$, respectively.
The modified Navier-Stokes with implicit free surface system \cref{momentumequation_nohydrostatic} and \cref{incompressibility2} at a time $n$ is therefore
\begin{equation}
\rho_0 \frac{\boldsymbol{u}^{n+1}- \boldsymbol{u}^n}{\Delta t}= R^{n+\theta} - \bnabla p^{n+\theta} - \rho^{n+\theta} g \boldsymbol{n}_g,
\label{momentumequation_nohydrostatic_discrete}
\end{equation}
\begin{equation}
\bnabla \bcdot \boldsymbol{u}^{n+1}=0,
\label{incompressibility_discrete}
\end{equation}
where $R^{n+\theta} = \theta R^{n+1} + (1-\theta) R^n$
contains the advective mass flux term,
together with viscosity and other source terms.
A choice $\theta \in [\frac{1}{2},1]$ leads to an implicit time stepping scheme that allows simulations to use large time steps, which are not restricted by the
Courant-Friedrichs-Lewy (CFL) condition~\cite{courant28}
with respect to the velocity and wave speed.
In practice for the simulations presented here, for the required level of accuracy and stability, Courant numbers up to 10 are applied.

\subsection{Combined free surface~--~pressure Chorin corrector}
\label{sec:corrector}
Under a Galerkin FE spatial discretisation
the
temporally discretised
momentum~\cref{momentumequation_nohydrostatic_discrete}
and continuity~\cref{incompressibility_discrete}
equations
are tested with
the velocity $\boldsymbol{\phi}_i$
and pressure $\psi_i$ basis functions, respectively.
The trial functions
$\boldsymbol{u}$
and
$p$
are defined in terms of their respective basis functions also, and
\cref{sec:spatial} describes their form and the nomenclature used in more detail.
This leads to the space-time discrete momentum equation
\begin{align}
& \rho_0 \frac{M_u}{\Delta t}
\left(
\boldsymbol{u}_*^{n+1}
-
\boldsymbol{u}^{n}
\right)
+
\theta \tilde{A}^{n+1}
\boldsymbol{u}_*^{n+1}
+
(1 - \theta) A^{n} \boldsymbol{u}^n
\nonumber
\\
&
\qquad
\qquad
\quad
= \theta C p_*^{n+1}
+(1-\theta) C p^n + S_u,
\label{eq1}
\\
&
\rho_0 \frac{M_u}{\Delta t}
\left(
\boldsymbol{u}^{n+1}
-
\boldsymbol{u}^{n}
\right)
+
\theta A^{n+1} \boldsymbol{u}^{n+1}
+
(1 - \theta) A^{n} \boldsymbol{u}^n
\nonumber
\\
&
\qquad
\qquad
\quad
= \theta C p^{n+1}
+(1-\theta) C p^n + S_u,
\label{eq2}
\end{align}
for a Picard iteration step and end of time step, respectively.
The starred variable
$\boldsymbol{u}_*^{n+1}$ is the current best approximation to $\boldsymbol{u}^{n+1}$, calculated from
pressure at the previous time level $n$.
The best guess of the solenoidal velocity
at a time level $n\!+\!1$
is denoted
$\tilde{\boldsymbol{u}}^{n+1}$,
and used in the calculation of updated non-linear operators,
such as mass flux $\tilde{A}^{n+1}$.
The velocity space mass matrix $M_u$
additionally contains
the diagonal or block-diagonal (depending on the chosen discretisation)
component of
viscosity from $R$, which is to be treated implicitly in pressure.
The discrete cross-space gradient operator
$C_{ij}\!:=-\int_\Omega \boldsymbol{\phi}_i \bnabla \psi_j \ud \Omega$,
contains an inner product over velocity and pressure spaces,
leaving sources in $S_u$.

Subtracting \cref{eq1} from \cref{eq2}
and multiplying by
\scalebox{0.95}[1]{$%
\theta{\Delta t}C^{T}\!M_u^{\!-\!1}$}
yields a discrete Poisson equation for the correction
\begin{equation}
\rho_0
\theta C^{T}\!(
\boldsymbol{u}^{n+1}
\!-
\boldsymbol{u}_*^{n+1}
)
=
\theta^2 \Delta t \, C^{T}\!M_u^{-1}
C (p^{n+1}
\!-
p_*^{n+1}).
\label{eq3}
\end{equation}

\subsection{Discrete continuity}
Discretisation of the continuity equation~\cref{incompressibility_discrete}
is written
\begin{equation}
\theta C^T \boldsymbol{u}^{n+1}+(1-\theta)C^T \boldsymbol{u}^{n}
 +G_{{\theta}}^T \boldsymbol{u}^{n+1} +G_{{1-\theta}}^T \boldsymbol{u}^{n} = 0,
\label{Ceq2}
\end{equation}
where
$G_{\theta, ij}\!:=\int_{\Gamma_\eta}\theta \boldsymbol{n} \boldsymbol{\phi}_i \psi_j \ud \Gamma$ and
$G_{(1-\theta), ij}\!:=\int_{\Gamma_\eta}(1-\theta) \boldsymbol{n} \boldsymbol{\phi}_i \psi_j \ud \Gamma$.
For $G_{\theta}=G_{(1-\theta)}=0$, the system
of equations enforces incompressibility with weakly applied no normal flow
boundary conditions.

\subsection{Discrete modified kinematic boundary condition}
Discretisation of the free surface boundary condition
\cref{FSeq1-1b} is now required to
provide the boundary integral terms in
\cref{Ceq2},
with a $\theta$ time discretisation described by
\begin{align}
&
\boldsymbol{n}^{n+1}
\bcdot
\boldsymbol{n}_g
p^{n+1}
-
\boldsymbol{n}^{n}
\bcdot
\boldsymbol{n}_g
p^{n}
\nonumber
\\
&
\qquad
=
\rho_0
g
\Delta t \,
\left(
\theta
\boldsymbol{n}^{n+1}
\bcdot
\boldsymbol{u}^{n+1}
+
(1-\theta)
\boldsymbol{n}^{n}
\bcdot
\boldsymbol{u}^{n}
\right).
\label{FSeq1-1b-time}
\end{align}
Discretisation of \cref{FSeq1-1b-time} in space using the test and trial functions
$\phi_i$ and $\psi_i$,
introduced in
\cref{sec:corrector} gives
\begin{equation}
M_s\frac{p^{n+1}
-
p^n} {\rho_0 g \Delta t}
=
G_\theta^T \boldsymbol{u}^{n+1}
+
G_{1-\theta}^T \boldsymbol{u}^{n},
\label{FSeq2}
\end{equation}
with the surface integral
$M_{s, ij}\!:=\int_{\Gamma_\eta}n_g \psi_i \psi_j \ud \Gamma$.

Applying this discrete combined \fp kinematic condition \cref{FSeq2} to the discrete continuity \cref{Ceq2}, we find
\begin{equation}
\theta C^T \boldsymbol{u}^{n+1}+(1-\theta)C^T \boldsymbol{u}^{n}
+M_s\frac{p^{n+1}-p^n}{\rho_0 g \Delta t} = 0.
\label{Ceq3}
\end{equation}
Substituting \cref{Ceq3} into the momentum equation \cref{eq3},
with the correction defined $\Delta p\!:= p^{n+1} - p_*^{n+1}$,
yields
\begin{align}
&\left(\theta^2 C^{T}M_u^{-1} C
+ \frac{M_s}{g (\Delta t)^2}\right) \Delta p
\nonumber\\ &
=
-\frac{
\theta C^T \boldsymbol{u}^{n+1}_*
\!+\!
(1\!-\!\theta)C^T \boldsymbol{u}^{n}
}{\Delta t}
-
\frac{M_s}{g (\Delta t)^2}(p_*^{n+1}\!-p^{n}).
\label{Ceq5}
\end{align}

\subsection{Combined free surface~--~pressure system solution}
\label{sec:solutionsummary}
During a single Picard iteration, the first velocity \emph{predictor} step solves
the discrete linearised momentum equation \cref{eq1}, to establish an updated
intermediate velocity
$\boldsymbol{u}_*^{n+1}$, from the current best approximation to the velocity and pressure, and their value at the previous time step.

The predictor $\boldsymbol{u}_*^{n+1}$ obtained is not divergence free in general, and in order to
enforce the incompressibility condition, a
\emph{pressure correction}
$\Delta p$
is calculated to project this velocity into
the divergence free subspace by solving
\cref{Ceq5} above.
The \emph{velocity correction} is made to update the intermediate velocity, consistent with the new intermediate pressure,
and projected to the divergence free subspace using the difference of \cref{eq1,eq2}, where
\begin{equation}
\boldsymbol{u}^{n+1}
=
\boldsymbol{u}^{n+1}_*
+
\frac{\theta \Delta t}{\rho_0}
M_u^{-1}
C
\Delta p.
\label{velocitycorrection}
\end{equation}

Finally, the \emph{interface tracking} step adjusts the
free surface position
following
\cref{pressurebc}
in light of the new pressure field, in a direction $-\boldsymbol{n}_g$, parallel to the gravitational vector.

\section{High aspect ratio wetting and drying domains}
\label{sec:conditioning}
\subsection{Wetting and drying of the simulation domain}
The free surface boundary
is split into
distinct
wet
and
dry
regions
(illustrated in \cref{fig:domain}),
defined by the combined \fp at the surface
such that $\Gamma_\eta = \Gamma_w \cup \Gamma_d$,
with
\begin{alignat}{3}
&
\Gamma_w\!:
\boldsymbol{r} \in \Gamma_\eta,
\quad
&&
\forall
\;
p(\boldsymbol{r}) \ge \rho_0 g (h(\boldsymbol{r}) + d_0),
\; \textrm{and}
\nonumber
\\
&
\Gamma_d\!:
\boldsymbol{r} \in \Gamma_\eta,
&&
\forall
\;
p(\boldsymbol{r}) < \rho_0 g (h(\boldsymbol{r}) + d_0).
\nonumber
\end{alignat}
The conditions in dry regions differ from those in wet in two defining ways.
Firstly, the water column depth is maintained at a threshold minimum $d_0$ above the bottom bathymetry defined by
$h\!: \Omega \mapsto \mathbb{R}$,
and secondly, the surface boundary condition on the combined \fp prognostic variable is modified to enforce this constraint in the solver.
With the depth constraint the free surface evolution described by
\cref{pressurebc} provides the first constraint
\begin{equation}
\eta(\boldsymbol{r}) = \textrm{max}
\left(
\frac{1}{\rho_0 g}
p(\boldsymbol{r}),
h(\boldsymbol{r}) + d_0
\right),
\;\;
\textrm{for} \;
\boldsymbol{r}
\;
\textrm{on}\;
\Gamma_\eta.
\label{wetdryconstraint}
\end{equation}
The second is found by modifying the combined kinematic condition \cref{FSeq1-1b}, which
in light of the depth restriction gives
\begin{equation}
\boldsymbol{n} \bcdot \boldsymbol{n}_g
\frac{\partial }{\partial t}
\textrm{max}
\left(
p, \rho_0 g \left( h + d_0 \right)
\right)
=
\rho_0 g \,
\boldsymbol{n} \bcdot \boldsymbol{u},
\;\; \textrm{on} \; \Gamma_\eta.
\label{FSeq1-1b-wetdry}
\end{equation}

In wet regions $\Gamma_w$,
the constraints
\cref{wetdryconstraint}
and
\cref{FSeq1-1b-wetdry}
reduce back to
the free surface conditions
\cref{pressurebc}
and
\cref{FSeq1-1b},
respectively.
In dry regions $\Gamma_w$,
\cref{FSeq1-1b-wetdry}
is a no normal flow condition,
and effectively imposes a rigid lid approximation.

\subsection{Conditioning of the pressure calculation}
\label{sec:conditioningofpressuresolve}
The spatial domains of geophysical processes are typically large aspect ratio, due to the gravitational influence and disparity in dynamics parallel and perpendicular to geoid surfaces.
The solution of a non-linear fluid flow system in these types of domains including non-hydrostatic dynamics,
with
implicit time evolution
and a predictor~--~corrector approach such as
\cref{sec:solutionsummary}
is shown in
\cite{kramer10}
to lead to an ill-conditioned pressure system.
In the limit of large domain aspect ratio and long time steps, the system behaves approximately as though it has a rigid lid, where the free surface is fixed with $\eta = 0$ and
$\boldsymbol{u} \bcdot \boldsymbol{n} = 0$.

The dry regions introduced by the WD process
significantly exacerbate ill-conditioning,
since
a rigid lid condition is applied
directly
and
the region
contains elements with acutely large aspect ratios
due to their defining shallow water column depth.

The correction
$\boldsymbol{u}^{n+1} = \boldsymbol{u}_* - \bnabla \phi$
\cref{velocitycorrection}
is calculated subject to the divergence-free constraint
$\bnabla \bcdot \boldsymbol{u}^{n+1} = 0$.
This leads to
the following pressure Poisson equation for $\phi$
\begin{equation}
\bnabla^2 \phi
=
\bnabla \bcdot \boldsymbol{u}_*,
\label{ppe}
\end{equation}
which corresponds to the discrete Poisson operator
$C^T M_u^{-1} C$ in the formulations \cref{Ceq5} and \cref{Ceq5-wetdry} above.
For no normal flow boundary conditions
where
$\boldsymbol{u} \bcdot \boldsymbol{n} = 0$,
at interfaces with bedrock or in the case of the rigid lid approximation for the ocean-air interface,
the coupling between velocity and pressure
results in
the corresponding boundary condition on
\cref{ppe} as the Neumann expression
\begin{equation}
\frac{\partial \phi}{\partial n}
= 0,
\;\;
\textrm{on}
\;
\Gamma_\eta,
\label{rigidlidbc}
\end{equation}
ensuring the velocity constraint is consistently preserved.

Applying a kinematic condition instead leads to the homogeneous Dirichlet condition $\phi = 0$ on $\Gamma_\eta$.
The redefinition of pressure in \cref{piezometric} to form the piezometric pressure here allows standard pressure splitting approaches to treat baroclinic and barotropic dynamics
\citep[e.g.][]{shchepetkin05} in the general case of domains discretised with fully-unstructured meshes.
These schemes themselves aid the conditioning of pressure solves in geophysical models \cite{maddison11}, where there is a large disparity of scales and resolution of the dominant physical processes.
This piezometric variable satisfies the same equation~\cref{ppe}, with a modified right-hand side source term and boundary condition.

Discretisation of the kinematic condition \cref{FSeq1-1b} defined in terms of the piezometric pressure using implicit backward Euler in time gives a Robin condition for $\phi$, such that
\begin{equation}
\boldsymbol{n} \bcdot \boldsymbol{n}_g
\frac{\phi}{{\Delta t}^2}
=
g
\frac{\partial \phi}{\partial n}.
\label{rigidlidbc2}
\end{equation}
With the barotropic wave speed $c = \sqrt{gH}$, for a distance $H$, and noting that
\begin{equation}
\frac{\phi}{{\Delta t}^2}
\Big/
g\frac{\partial \phi}{\partial n}
\approx
\frac{H}
{g{\Delta t}^2}
=
\left(
\frac{H}{c\Delta t}
\right)^2,
\nonumber
\end{equation}
the ratio of the terms in \cref{rigidlidbc2} scale as
the square of the time it takes for a barotropic wave to travel a distance $H$ relative to the length of a time step, and we see that the condition for free surface flows \cref{rigidlidbc2} tends to that of the rigid lid \cref{rigidlidbc} in the large time step limit.
So although adjusting a system to apply a free surface kinematic boundary condition on the top surface as opposed to a rigid lid does improve conditioning for modest aspect ratios, as the disparity in scales increases and the aspect ratio becomes smaller, or equivalently
larger time steps are taken, the ill-conditioning of a rigid lid system is soon recovered due to the quadratic dependence.

The multigrid preconditioner of \cite{kramer10} for unstructured meshes on high aspect ratio domains helps better condition the Poisson problem
in general, without consideration of WD,
using a combination of algebraic multigrid and a geometric vertical prolongation operator.
This solver method itself makes it feasible to run non-hydrostatic unstructured mesh simulations of fluids in geophysical domains.

Whilst the relatively moderate aspect ratio wet areas can be treated by the multigrid preconditioner approach,
specific methods to handle the acute aspect ratio and direct rigid lid condition applied in dry regions are required,
if this general fully \threed and non-hydrostatic WD approach is to be
applied to real geophysical systems.

\subsection{Quantification of the ill-conditioning}
\label{sec:verticalrelax}
The discrete form of the Laplacian operator that appears on the left-hand side of \cref{ppe},
seen in
\cref{Ceq5},
has eigenvalues
$\lambda_i \sim k_i^2$, for wavenumbers $k_i$.
The conditioning of the matrix
is determined by the ratio of the
maximum
$\norm{ \lambda }_\infty$
and
minimum
$\norm{ \lambda }_\mathrm{min}$
eigenvalues.
This is, equivalently, the ratio of the minimum and maximum wavenumbers,
$k_\mathrm{min}$ and $k_\mathrm{max}$,
squared
\begin{equation}
\kappa
\left(
C^{T}\!M_u^{-1} C
\right)
=
\abs{
\frac{ \norm{ \lambda }_\infty }
 { \norm{ \lambda }_\mathrm{min} }
}
\sim
\abs{
\frac{ \norm{ k^2 }_\infty }
 { \norm{ k^2 }_\mathrm{min} }
}
=
\abs{
\frac{ \norm{ k }_\infty }
 { \norm{ k }_\mathrm{min} }
}^2\!\!.
\end{equation}
For high aspect ratio problems
$H / L \ll 1$, for $H$ and $L$ characteristic length scales of the solution domain in the vertical and horizontal, respectively
(see \cref{fig:domain}), we find
\begin{align}
&\norm{ k }_\mathrm{min} \sim \frac{1}{H},
\;\;
\textrm{and}
\;\;
\norm{ k }_\infty
\sim
\frac{1}{L},\ \textrm{and hence}
\nonumber
\\
&
\qquad
\kappa
\left(
C^T M_u^{-1} C
\right)
\sim
\left(
\frac{ H }
 { L }
\right)^{2}.
\label{conditionscaling}
\end{align}
On a spheroid, such as the Earth, the characteristic `horizontal' length scale $L$ is the extent of the encompassing surface geoid,
with $H$ the height in a direction parallel to gravitational acceleration.
Conditioning of linear system that results from the discretisation of the Poisson equation is approximately proportional to the square of the aspect ratio of the global domain.
Equivalently, the element edge-lengths, which are constrained to resolve processes important to the simulation, can also be used to characterise the scaling, such that condition number is proportional to
${(\Delta x / \Delta z})^2$,
with $\Delta x$ and $\Delta z$ characteristic edge-lengths in local horizontal and vertical directions, respectively.

Entries into the matrix of the linear system that arise from dry cells would ideally be removed, in a process similar to lifted Dirichlet boundary conditions \cite[e.g.][]{karniadakis05} and the solver limited to variables on the wet sub-system, much like an element removal approach.
For an implicit approach it is not clear how this would be accomplished without adversely affecting the natural evolution of the interface.
Instead, under implicit integration, treatment of the ill-conditioning
highlighted by~\cref{conditionscaling} needs to be addressed.

\subsection{Vertical velocity relaxation in dry areas}
To close the system
\cref{momentumequation_nohydrostatic}--\cref{incompressibility2},
an equation of state is required.
Details of the form of this function do not influence the development that follows,
and a general treatment of
the evolution of density is considered, such that
\begin{equation}
\frac{\partial \rho}{\partial t}
+
\boldsymbol{u} \bcdot \bnabla \rho = 0,
\label{d2eqnstrong}
\end{equation}
with its temporal discretisation following \cref{sec:temporal} as
\begin{equation}
\frac{\rho^{n+1} -\rho^n}{\Delta t} + \tilde{\boldsymbol{u}}^{n+1}\bcdot \bnabla \rho^{n+1} = 0.
\label{d2eqnnew}
\end{equation}
Development of the approach proceeds with a
discretisation of the density transport equation~\cref{d2eqnstrong},
in a slightly different linearisation to that of~\cref{d2eqnnew}, of the form
\begin{align}
\frac{\rho^{n+1} -\rho^n}{\Delta t}
+ w^{n+1} \frac{\partial \rho_*^{n+1}}{\partial z} + {s}_{\rho*}^{n+1}=0,
\label{d2eqn-sim-a}
\\
\frac{\rho^{n+1}_* -\rho^n}{\Delta t}
+ w_*^{n+1} \frac{\partial \rho_*^{n+1}}{\partial z} + {s}_{\rho*}^{n+1}=0,
\label{d2eqn-sim-b}
\end{align}
with vertical velocity $w$,
starred variables representing the best current guess, and
the source term
$
s_{\rho}^{n+1}
$
containing details of spatial gradients of density locally aligned to the geoid.
Subtracting
\cref{d2eqn-sim-b}
from
\cref{d2eqn-sim-a}
gives a transport equation
that mirrors \cref{eq3},
describing the variation over the Picard iteration process
\begin{equation}
\frac{\rho^{n+1} -\rho_*^{n+1}}{\Delta t}
+
(w^{n+1} - w^{n+1}_*)
\frac{\partial \rho_*^{n+1}}{\partial z}
= 0.
\label{d2eqn-sim}
\end{equation}
Substitution of this temporally discrete density transport equation~\cref{d2eqn-sim} into the momentum equation~\cref{momentumequation_nohydrostatic_discrete}
leads to
\begin{align}
&\rho_0 \frac{\boldsymbol{u}^{n+1}- \boldsymbol{u}^n}{\Delta t}
=
R^{n+\theta}
- \bnabla p^{n+\theta}
\nonumber
\\
&
\quad
+ g \boldsymbol{n}_g \Delta t \frac{\partial {\rho}_*^{n+1}}{\partial z} (w^{n+1}- {w}_*^{n+1})
- {\rho}_*^{n+1} g \boldsymbol{n}_g.
\label{ns2}
\end{align}

The FE weak form of \cref{ns2} is developed by testing
with velocity basis functions
$\boldsymbol{\phi}_i$ and applying integration by parts twice at the free surface
to obtain
\begin{align}
& \int_\Omega
\boldsymbol{\phi}_i
\rho_0
\frac{\boldsymbol{u}^{n+1} - \boldsymbol{u}^n}{\Delta t} \ud \Omega
=
\int_\Omega \boldsymbol{\phi}_i
\Big(
R^{n+\theta}
- \bnabla p^{n+\theta}
\nonumber
\\
&
\;\;
+
g \boldsymbol{n}_g
\Delta t \frac{\partial \rho_*^{n+1}}{\partial z} (w^{n+1}- w_*^{n+1})
-
\rho_*^{n+1}
g \boldsymbol{n}_g
\Big)
\ud \Omega
\nonumber
\\
&
\;\;
- \int_{\Gamma_\eta} \boldsymbol{\phi}_i \boldsymbol{n} \bcdot \boldsymbol{n}_g g \Delta t (\rho^{n+1}-\rho_{a})(w^{n+1} - w_*^{n+1}) \ud \Gamma.
\label{ns-dis}
\end{align}
The density of air just above the free surface interface $\rho_{a}$, can in most cases be neglected as a small effect, in the same way as the atmospheric pressure.

Assuming that, for shallow waters, the vertical velocity $w$
is linearly related to the distance from the bottom of the ocean
or in a depth-averaged sense
and ignoring the density variations $\rho'$
in the surface integral above,
the terms in \cref{ns-dis} above containing explicit reference to the vertical velocity $w$
can be grouped into an absorption term
\begin{equation}
\sigma_{zz}= g \Delta t
\,
\max\left(
0,
- \frac{\partial \rho_*^{n+1}}{\partial z}
\right)
+
\frac{
g \Delta t
\rho_0
}
{d}
\boldsymbol{n} \bcdot \boldsymbol{n}_g
,
\label{sigma_z}
\end{equation}
where $d = h + \eta$ is the water depth,
such that
\begin{align}
&\int_\Omega \boldsymbol{\phi}_i \rho_0 \frac{\boldsymbol{u}^{n+1} - \boldsymbol{u}^n}{\Delta t} \ud \Omega
=
\int_\Omega \boldsymbol{\phi}_i
\Big(
R^{n+\theta}
- \bnabla p^{n+\theta}
\nonumber
\qquad\qquad
\\
&
\qquad
-
\underbrace{
\sigma (\boldsymbol{u}^{n+1} - \boldsymbol{u}_*^{n+1})
}_\dagger
- \rho_*^{n+1} g \boldsymbol{n}_g
\Big)
\ud \Omega,
\nonumber
\\[-12pt]
&
\qquad
\qquad
\qquad
\qquad
\qquad
\textrm{with}
\quad
\sigma =
 \begin{pmatrix}
 0 & 0 & 0 \\
 0 & 0 & 0 \\
 0 & 0 & \sigma_{zz}
 \end{pmatrix}.
\label{ns-dis-2}
\end{align}
The inverse time scale for the vertical velocity
relaxation is defined by \cref{sigma_z}.
As the Picard iterations proceed
and
$
\boldsymbol{u}_*^{n+1}
\to
\boldsymbol{u}^{n+1}
$,
the magnitude of this stabilising term, marked by
$\dagger$
in
\cref{ns-dis-2},
relaxes to zero.
Although the absorption coefficient $\sigma$ will be relatively small in wet regions, and the contribution from $\dagger$ small overall, it is important to include these terms throughout in order to maintain consistency and as a result, accuracy.

The following conditions on
vertical density gradient,
the free surface,
vertical viscosity,
and
vertical absorption
provide a well-conditioned
pressure Poisson equation
\begin{enumerate}
\item \emph{Vertical density gradient}
\begin{equation}
\frac{(\Delta x)^2}{(\Delta z)^2}
\leq
a^2
g
{\Delta t}^2
\textrm{max}
\left(
\frac{\partial \rho_*^{n+1}}{\partial z}, 0
\right),
\end{equation}
\item \emph{Free surface variation}
\begin{equation}
\frac{(\Delta x)^2}{(\Delta z)^2} \leq \frac{ a^2 (\Delta t)^2 g }{d},
\label{vert-fs}
\end{equation}
\item \emph{Vertical viscosity}
\begin{equation}
\frac{(\Delta x)^2}{(\Delta z)^2} \leq \frac{ a^2 \Delta t \, \nu_{zz} }{\Delta z^2},
\label{vert-vis}
\end{equation}
\item \emph{Vertical absorption}
\begin{equation}
\frac{(\Delta x)^2}{(\Delta z)^2} \leq a^2 \Delta t \, \sigma_{zz},
\label{vert-abs}
\end{equation}
\end{enumerate}
where $a$ is a tolerable aspect ratio of element length scales
(e.g. unity in the isotropic case),
$\Delta x$
and
$\Delta z$
characterise local resolution scales,
$\nu_{zz}$
is a kinematic viscosity,
and
$\sigma_{zz}$
an absorption.
Note that the viscosity of \cref{vert-vis}
must be treated implicitly
or semi-implicitly
in pressure
(e.g. diagonal or block diagonal)
in order to control the condition
number of the pressure Laplacian.
Implementation of the viscosity in stress form is appropriate here
since tensor forms directly smooth horizontal velocities in the vertical.

In the case of WD,
where \cref{vert-fs} does not hold,
we must ensure \cref{vert-abs} is satisfied
by a suitable
choice of the absorption $\sigma_{zz}$.
From \cref{vert-abs}, in order to make the resulting pressure matrix feel like an $O(a)$
aspect ratio domain, we need
\begin{equation}
\sigma_{zz}
=
\frac{(\Delta x)^2}{a^2 \Delta t (\Delta z)^2}.
\label{sigma_z_for_order_one}
\end{equation}
The form of $\sigma_{zz}$
in
\cref{sigma_z_for_order_one}
defines the inverse time scale for the vertical velocity
relaxation in \cref{sigma_z}.
Note that this form of $\sigma_{zz}$ is spatially varying, and in particular the characteristic local length scales
$\Delta x$
and
$\Delta z$
are non-homogeneous across the geoid surface.
In WD simulations these fields contain large deviations which are indicative of the regions affecting conditioning of the pressure Poisson equation.

\subsection{Discretisation for high aspect ratio domains}
The momentum equation
\cref{ns-dis-2}
discretised in
space-time at any given Picard iteration step is
\begin{align}
& \rho_0 \frac{M_u}{\Delta t}
\left(
\boldsymbol{u}_*^{n+1}
-
\boldsymbol{u}^{n}
\right)
+
\theta \tilde{A}^{n+1}
\boldsymbol{u}_*^{n+1}
+
(1 - \theta) A^{n} \boldsymbol{u}^n
\nonumber
\\
&
\ =
\theta C p_*^{n+1}
+(1-\theta) C p^n
-
Q
\left(
\boldsymbol{u}^{n+1}
\!-
\boldsymbol{u}_*^{n+1}
\right)
+ S_u,
\label{eq1mod}
\end{align}
with
$Q_{ij}\!:=\!\int_\Omega \phi_i \sigma \phi_j \ud \Gamma$.
This balance
compared to its end of time step state
is
multiplied by
$\theta{\Delta t}C^{T}\!M_u^{-1}$,
to give
\begin{align}
&
\left(
\rho_0
\theta C^T
-
\theta \Delta t \, C^T
M_u^{-1}
Q
\right)
\left(
\boldsymbol{u}^{n+1}
-
\boldsymbol{u}_*^{n+1}
\right)
\nonumber
\\
&
\qquad
\qquad
\qquad
=
\theta^2 \Delta t \, C^T
M_u^{-1}
C (p^{n+1}-p_*^{n+1})
\label{eq3mod}.
\end{align}
This is equivalent to
\cref{eq3} previously, noting that the term marked $\dagger$
in \cref{ns-dis-2}
is zero at the end of a time step.

The new form of the combined kinematic boundary condition
\cref{FSeq1-1b-wetdry} leads to a time discretised form modified from
\cref{FSeq1-1b-time} to include the no normal flow component applied in dry regions and is described by
\begin{align}
& \boldsymbol{n}^{n+1}
\bcdot
\boldsymbol{n}_g
\mathrm{max}
\left(
p^{n+1},
\rho_0 g
\left(
h+d_0
\right)
\right)
\nonumber
\\
&
\qquad
-
\boldsymbol{n}^{n}
\bcdot
\boldsymbol{n}_g
\mathrm{max}
\left(
p^{n},
\rho_0 g
\left(
h+d_0
\right)
\right)
\nonumber
\\
&
\qquad
\quad
=
\rho_0
g
\Delta t \,
\left(
\theta
\boldsymbol{n}^{n+1}
\bcdot
\boldsymbol{u}^{n+1}
+
(1-\theta)
\boldsymbol{n}^{n}
\bcdot
\boldsymbol{u}^{n}
\right).
\nonumber
\end{align}
Moreover, the surface integral
$M_s$
is modified such that
the discrete modified kinematic condition \cref{FSeq2} becomes
\begin{equation}
M_{w}\frac{p^{n+1}
-
p^n} {\rho_0 g \Delta t}
+
M_{d}
\frac%
{h + d_0}
{\Delta t}
=
G_\theta^T \boldsymbol{u}^{n+1}
+
G_{1-\theta}^T \boldsymbol{u}^{n},
\nonumber
\end{equation}
where
$M_{w, ij}=\int_{\Gamma_w}n_g \psi_i \psi_j \ud \Gamma$
and
$M_{d, ij}=\int_{\Gamma_d}n_g \psi_i \psi_j \ud \Gamma$.
This kinematic condition change modifies the pressure correction
and the discrete continuity \cref{Ceq3} becomes
\begin{equation}
\theta C^{T}\!\boldsymbol{u}^{n+1}
\!+
(1\!-\!\theta)C^{T}\!\boldsymbol{u}^{n}
\!+\!
M_{w}\frac{p^{n+1}
\!\!-\!
p^n} {\rho_0 g \Delta t}
+
M_{d}
\frac%
{h \!+\! d_0}
{\Delta t}
= 0.
\label{Ceq3-wetdry}
\end{equation}
Substituting \cref{Ceq3-wetdry} into momentum \cref{eq3mod} yields
the discrete
combined \fp Poisson corrector, with
\cref{Ceq5} evolving to
\begin{align}
&
\left(\theta^2 C^T
M_u^{-1} C
+ \frac{M_w}{g (\Delta t)^2}\right) \Delta p
\nonumber \\
&
\qquad
=
-\frac{
\theta C^T \boldsymbol{u}^{n+1}_*+(1-\theta)C^T \boldsymbol{u}^{n}
}{\Delta t}
-
\frac{
Q
\left(
\boldsymbol{u}_*^{n+1}
-
\boldsymbol{u}^{n+1}
\right)
}{\Delta t}
\nonumber \\
&
\qquad
\quad
-
\frac{M_w}{g (\Delta t)^2}(p_*^{n+1}-p^{n})
-
\rho_0
M_{d}
\frac%
{h + d_0}
{(\Delta t)^2}.
\label{Ceq5-wetdry}
\end{align}
The predictor~--~corrector method of \cref{sec:solutionsummary}
solves the non-linear system
with the updated \fp Poisson corrector \cref{Ceq5-wetdry}
combined with discrete linearised momentum
\cref{eq1mod},
and a velocity correction determined
from
\cref{eq3mod}.

\subsection{Self-consistency and physical basis of the solution}
Mass, momentum and tracer quantities are self-consistent and conserved, properties inherited from their underlying Galerkin FE weak formulations
\cite{piggott08}
and use of a thin-film
\cite{medeiros13,funke11}.
The constrained discrete Sobolev solution space of the weak form modified with the additional terms marked $\dagger$ in \cref{ns-dis-2}
converges on the solution space of the original form without these, as the
the Picard process proceeds.
In a similar manner to Petrov-Galerkin and variational multiscale \cite{hughes98} residual-based stabilisation methods
\cite{candy08},
this ensures consistency, that the solution found is a valid solution of the original weak Galerkin formulation, a true discrete solution to the governing continuum equations and is therefore physically-based.

Moreover, just like streamline-upwind Petrov-Galerkin (SUPG) stabilisation, the additional terms themselves are defined from physical properties of the flow.
For example, \eqref{sigma_z} includes contributions from
$g$,
the local vertical density gradient and water column depth.
This is supported along with local discretisation parameters such as time step and element size used to quantify unresolved scales,
in a similar way to multiscale turbulence closures \cite{candy08}.

\subsection{Determination of characteristic length scales}
\label{sec:lengthscales}
Accurate calculation of the characteristic length scales is critical to the success of the approach, particularly due to the quadratic dependence
in \cref{sigma_z_for_order_one}.

The calculation of the characteristic horizontal length scale could be simply the minimum or maximum edge length of the element projected to a \twod horizontal geoid.
A more accurate approximation can be determined from the smallest and largest circumscribing circular bounds of this projection.
The length scale $\Delta x$ is then a function of these minimum and maximum extents.
This is a natural approach for models employing anisotropic mesh elements.

The vertical length scale is less ambiguous to determine, since
unique intersections with $\Gamma_\eta$ and $\Gamma_b$ exist $\forall \boldsymbol{x}\in\Omega$,
due to the construction of geophysical domains \citep{candybrep,candy14},
and similarly for internal layers.
Evaluating length scale functions at Gaussian quadrature points rather than by element further increases accuracy, since FE assembly integrations are performed this way, with options to develop mean or area-weighted means.
This is trivially extended to superparametric elements which are typically used in the top layer for accurate representation of geoid curvature.

Arguably the best characterisation of tetrahedral element size is determined from the Jacobian transformation matrix which projects a FE from global to local parameterised space.
The determinant of the transformation Jacobian intersected with the local (to quadrature point) surface geoid plane and gravitational acceleration vector will give characteristic length scales for the element in the required horizontal and vertical directions, respectively.
This approach also naturally handles element anisotrophy and meshes which are fully unstructured in 3D.
The merits of this choice are examined in \cref{dztests}.

\subsection{Correction to velocity relaxation in shallow regions}
Under no forcing the momentum equation
\cref{ns-dis-2}
tends to relax the implicit velocity
$\boldsymbol{u}_*^{n+1}$
to the state in the previous time step
$\boldsymbol{u}^{n}$,
but this can be too strong in very shallow areas.
This is corrected by reducing the
magnitude of the explicit part of the velocity
that we relax to, by adding
$
-\gamma
\boldsymbol{u}^{n}
$
to the right of the momentum equation,
with
\begin{equation}
\gamma =
\mathrm{max}
\left(
2
\left(
1
-
\frac{d}{2d_0}
\right)\!,\  0
\right).
\end{equation}
For a water column depth $d$, this relaxation scales away the velocity in the vicinity of dry regions where $d < 2d_0$, and relaxes to zero in dry regions, where
$d = d_0$.

\section{Mesh movement with wetting and drying}
\label{sec:meshmovement}
\subsection{Discrete function space updates}
The free surface evolution results in many quantities varying in time, such as the free surface normal vector $\boldsymbol{n}$ in \cref{FSeq1-1b-time}.
Moreover, this includes the mesh, and hence spatial discretisation,
which leads to a change of the discrete function spaces $S^h$, and their spanning basis sets
resulting in
new forms of mass and other matrices in discrete forms such as \cref{FSeq2}.
For conservation and accuracy it is necessary to update the discrete non-linear system during the Picard iteration to reflect these changes.
There are various techniques to handle this conservatively, through the definition of a grid velocity, for example.
In this formulation, the domain discretisation is updated at the end of a Picard iteration to reflect the new free surface height predicted, with the normal $\boldsymbol{n}$, mass matrix and other matrices representing advection and surface integrals recalculated under the new domain discretisation.
It is therefore the case that the discrete matrices
$M_u$,
$M_s$,
$C$,
$A$,
$G$,
and
$Q$;
free surface normal $\boldsymbol{n}$,
basis functions $\boldsymbol{\phi}$ and $\psi$,
domain $\Omega$ and free surface $\Gamma_\eta$,
are always
the best known approximation,
i.e. the starred $n\!+\!1$ case.
A subtle exception is at the end of the final Picard iteration, where the update is not made, to ensure the domain and derivative parameters are those the prognostic variables were calculated on.

The generalised approach that includes the non-linear advection term in the governing equations \cref{momentumequation_nohydrostatic} precludes the discretised spatial operator from being self-adjoint.
Evaluation of this non-linear term requires sub-cycling, and for under-resolved high Froude number or rapidly-varying flows this could require a large number of iterations to converge,
unless the continuum system is linearised, or local resolution increased.

\subsection{Surface representation and interface tracking}
At the end of each Picard iteration,
as outlined in
\cref{sec:discretefreesurface},
the free surface position is updated using
\cref{wetdryconstraint}
to reflect the new pressure at the
interface
$p^{n+1}\big|_{\eta_s}$.
Due to the minimum threshold $d_0$,
the perturbation of the interface in the direction of the gravitational
acceleration is limited.
If the pressure $p^{n+1}$ at the interface implies it should move below this level,
it is fixed at the threshold level above the bottom bathymetry (i.e.
$\eta = h + d_0$).
The pressure remains unaffected, and is allowed to deviate from the
interface position $\eta$.
Conversely,
as soon as $p^{n+1}$ produces
a water column depth greater than $d_{0}$
the free surface interface moves upwards.
Correspondingly, the
domain discretisation is updated with the mesh stretched in the direction
$-\boldsymbol{n}_g$, parallel to the gravitational vector, to meet the new free surface bound.

Spatial representation of $\eta$ is inherited from
the function space used to approximate the combined \fp.
Irrespective of the order of variation, such as quadratic for the
\podgpt
element pair,
the interface
is approximated by a piecewise linear function as far as
the domain representation is concerned.
This satisfies the min-max property, such that the extent of the surface is bounded by the nodal positions that define its representation.
This, together with the
minimum threshold level prevents elements from becoming inverted
or excessively small.

\subsection{Remeshing}
It is not a requirement that the domain is remeshed anew to these adjusted bounds.
Since only one of the domain boundaries is perturbed through the above process
and in a direction aligned to the gravitational vector field, locally orthogonal to other bounds of the domain,
it is possible to
apply a relatively simple $r$-adaptive transform.
The domain mesh is stretched linearly in this direction to fit the new boundary.
It is also possible to limit the perturbation to the nodes on the free surface, or to apply more complicated $r$- or $h$-adaptive strategies to achieve a hybridised coordinate
system~\citep[see][]{kleptsova2010,bleck2002,burchard1997}
for more accurate solutions or better-represented features.
The implementation of the approach described here in the model code
\citep[Fluidity,][]{piggott08} functions with and supports these methods.

\section{Additional stabilising approaches for dry areas in high aspect ratio domains}
\label{sec:additional}
Two supplementary approaches to control conditioning are presented,
acting directly to prevent strong erroneous flows developing in the thin film and modifying behaviour in neighbouring wet regions.
This is exacerbated by the fact the physical system is solved in a weak sense, which whilst better for conditioning can permit large fluxes across the interface.
Unlike the above, these approaches transform the solution space and
have the potential to affect the solution in unphysical ways.
They are presented as additional techniques which can be employed to enable a solution to be reached, but require careful application.

\subsection{Manning-Strickler drag and dry region stability}
\label{sec:big_drag}
In the case of inundation flows where WD is applied, a parameterisation of drag that is
commonly employed is the
Manning-Strickler formulation, defining the bottom stress
\begin{equation}
\boldsymbol{n} \bcdot \mu \bnabla \boldsymbol{u}
=
n^2 g \frac{\left|\boldsymbol{u}\right|\boldsymbol{u}}{d^{1/3}},
\;\;
\textrm{on} \
\Gamma_{b},
\label{manning}
\end{equation}
where $n$ is the Manning coefficient, $d$ is the water depth and
$\boldsymbol{n}$ here is the unit surface normal on the bottom surface $\Gamma_b$.
This formulation itself has a stabilising effect, and more so in the very shallow dry regions,
with a drag applied along the bottom boundary proportional to $d^{-1/3}$.
In practice, the Manning-Strickler bottom stress is sufficient to prevent significant erroneous flow developing in dry areas.
In the cases of acute high aspect ratio, long time steps or particularly steep bathymetric gradients, the Manning coefficient can be increased in dry regions and their proximity to increase the stabilising effect,
with
\begin{equation}
{\hat n}
=
n
+
\mathrm{max}%
\left(
0,
n_\mathrm{dry}
\frac{2d_{0}-d}{d_{0}}
\right),
\nonumber
\end{equation}
where ${\hat n}$ replaces $n$ in
\cref{manning}, and
for $n_\mathrm{dry}$ a new Manning coefficient
(with usual standard units of
$\textrm{sm}^{-1/3}$)
large in size, relative to the standard coefficient $n$.

\subsection{Horizontal bulk eddy viscosity in dry regions}
\label{sec:viscosityindry}
A second solution to increase stability, is to damp flow directly in dry regions with a bulk volume viscosity or a source-absorption sponge, both allowing the approach to remain implicit.

This stabilisation is applied throughout the domain, or selectively in dry regions and their immediate proximity,
with the
large horizontal viscosity
\begin{equation}
\nu_L
=
\mathrm{max}%
\left(
0,
\nu_\mathrm{dry}
\frac{2d_{0}-d}{d_{0}}
\right),
\label{big_v}
\end{equation}
introduced to control spurious horizontal fluxes,
with
$\nu_\mathrm{dry}$
a constant eddy viscosity coefficient
and
$d \ge d_0 \ \forall \boldsymbol{x} \in \Omega$.
This horizontal viscosity is continuous in space without
discontinuous jumps in intensity across the WD interface, acting in the proximity of dry regions where
$d < 2d_0$.

\section{Validation and application: Numerical tests}
\label{sec:application}
Performance of the
implicit WD formulation described in
\cref{sec:equations,sec:discetisation,sec:conditioning,sec:meshmovement},
and
additional strategies of \cref{sec:additional}
are examined
in four test scenarios
in acutely high aspect ratio domains, to a degree found in geophysical systems.

\subsection{Implementation and verification}
The approach has been implemented and validated in the FE fluid dynamics code Fluidity~\cite{piggott08}.
This simulation framework contains many tools for geophysical modelling, is parallelised with sophisticated load balancing and supports adaptive mesh methods allowing computational effort to be focused on regions of dynamic interest.
It functions for a spatially variable gravitational acceleration vector, and hence can be used for large-scale simulations on the Earth's spheroid.
The implementation includes a suite of test cases to routinely verify the new algorithm in a formal sense, in an automated continuous verification
build engine~\cite{farrell11} to ensure robustness of the code and resiliency in light of further development.
The unstructured meshes used in the following cases were built by means of the open source software
Gmsh\footnote{\url{http://www.geuz.org/gmsh}.}.

The balance and LBB stability properties of the
\podgpt
velocity-pressure pairing
\citep[see][]{cotter09}
aid conditioning and are used in all applications considered here.
All four cases have been run on the purely continuous pairing
\popo also, but due to the pressure filtering required, did not perform as well,
and in all but modest aspect ratio cases were too ill-conditioned to reach convergence.
The behaviour of \podgpt and \popo and their relative performance in regular aspect ratio problems is presented in \cite{cotter09}.

Due to the aspect ratios considered, all cases use the multigrid preconditioner described in \cite{kramer10}
for iterative solution of the conditioned symmetric pressure Poisson linear system
in combination with Conjugate Gradient \citep[CG,][]{hestenes52}.
The momentum system is solved in a more standard approach with Symmetric Successive Over-Relaxation~\cite[SSOR,][]{young71} preconditioning and the iterative Restarted Generalised Minimal Residual~\cite[GMRES,][]{saad86} algorithm,
where the calculation is restarted after $k=30$ iterations.
The iterative SSOR-GMRES process is performed using algorithms built into the established and well-verified PETSc library~\cite{petsc}.
In contrast to the study \cite{funke11}, it was found that two Picard iterations provide sufficient convergence of the coupled system in the cases studied.
In all cases, both linear systems are solved to a convergence criteria specified by a relative error tolerance of ${10}^{-7}$, which is considered sufficiently accurate.
The quadrature based subgrid resolution described in~\cite{funke11} is also used, with a quadrature degree of eight.

\subsection{First Balzano sloped channel benchmark}
\label{sec:balzano1}
\begin{SCfigure*}
\centering
\includegraphics[width=13.9cm]{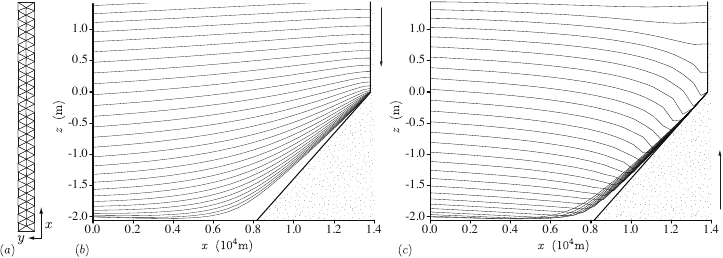}
\vspace{-0.3cm}
\caption{%
The first Balzano channel flow benchmark with, in this presented case, a horizontal extent of $1.38\times{10}^4\textrm{m}$, corresponding to a minimum element aspect ratio of ${\sim{10}^{-6}}$.
The discretised horizontal surface (a)
contains 108 triangular elements
with a characteristic length scale of ${\sim500\textrm{m}}$.
Vertical sections show the free surface position at 10min intervals
for the initial drying phase (b) and during wetting (c).
}
\label{fig:balzano1}
\end{SCfigure*}
\begin{figure}[!ht]
\centering
\includegraphics[width=8.3cm]{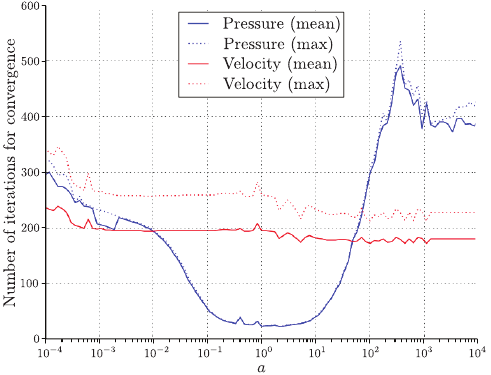}
\vspace{-0.3cm}
\caption{%
Impact of the optimal aspect ratio parameter on solver conditioning in the first Balzano benchmark
over a WD phase. 101 individual simulations shown.
}
\label{fig:balzano_a}
\end{figure}
\begin{figure}[!ht]
\centering
\includegraphics[width=8.3cm]{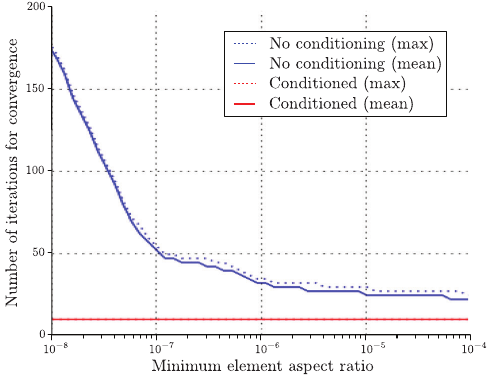}
\vspace{-0.3cm}
\caption{%
Pressure solver iterations to convergence in the first Balzano benchmark with respect to domain global aspect ratio,
in 101 individual simulations,
with and without conditioning applied.
}
\label{fig:balzano_its}
\end{figure}

The first two sets of numerical tests are from the suite of problems in Balzano \cite{balzano98},
selected since they exhibit the problematic ill-conditioning in as simple a setup as possible.
No analytical solution is available, so the problem configuration is chosen consistently with \cite{balzano98} to be able to draw
comparisons.
The base benchmark case is developed
from the originally \twod domain consisting of a 13.8km long slope with a depth of 5m at one end which tends to zero
at the other.
Recently developed schemes, such as the
flux-limiting WD method for FE SWE models presented in \cite{gourgue09}
and
the non-hydrostatic algorithm proposed in \cite{funke11},
have been benchmarked on these cases.
These model in \threed,
but force dynamics to occur predominantly in the directions where the extremes in extent occur,
with 10 elements introduced in the third direction in the former and 1-2 in the latter,
which is followed here to a width of $1\textrm{km}$.
With the assumption solutions are laminar, this extrusion into \threed space will not change the physical behaviour.
The sloped bottom bathymetry is defined
$h(x) = x / 2760$,
for the $x$-coordinate direction indicated alongside
the surface geoid computational mesh in \cref{fig:balzano1}(a).
The base case single-layer mesh contains vertically-aligned nodes and a horizontal element size of 500m.

Following the benchmark description in
Balzano \cite{balzano98} (also in \cite{gourgue09}),
no normal flow boundary conditions are applied at the bottom and shallow end of the domain,
and additionally applied to the sides.
A Manning-Strickler drag with $n\!=\!0.02\textrm{sm}^{-1/3}$ is applied at the bottom boundary.
The gravitational acceleration is set to $9.81\mathrm{m}\mathrm{s}^{-2}$ and the fluid is initially at rest.
Time discretisation is performed with Crank-Nicholson integration (i.e. $\theta\!=\!\frac{1}{2}$) and a time step of 600s.
In this case the WD threshold is set at $d_0\!=\!0.5\textrm{mm}$.
The free surface is forced at the deep open boundary with
a sinusoidal variation
of amplitude 2m,
such that water column thickness oscillates between 3--7m,
with a period of
12h.

In the series of tests considered here, the horizontal extent is varied from
$1.38\times{10}^2\textrm{m}$
to
$1.38\times{10}^6\textrm{m}$,
centred about the defined benchmark length of
$1.38\times{10}^4\textrm{m}$.
This provides a range of element aspect ratios from
${10}^{-4}$
to
${10}^{-8}$,
a domain aspect ratio up to
$3.62\times{10}^{-6}$
and spatial scales spanning over 10 orders of magnitude in a single domain.
Element lengths are scaled with the domain length, such that element aspect ratio relative to global aspect ratio is maintained,
with the extrusion in the third direction also scaled to preserve element shape.
The time step is also scaled to ensure the wave Courant number is constant.
The WD threshold $d_0$, and vertical extent are kept constant across all cases.

The free surface evolution of the intermediate case with a horizontal extent of
$1.38\times{10}^4\textrm{m}$
is shown in \cref{fig:balzano1} at 10min intervals, matching \cite{balzano98} and \cite{gourgue09}, for the initial drying and then wetting phase, respectively.
The results are physically reasonable and comparable to other formulations
(\cite{funke11} and \cite{gourgue09} for example).
In particular, the free surface interface suffers from neither underestimation with negative water column thickness, nor does it produce oscillations during the wetting process observed in \cite{balzano98} for some of the 10 methods examined.
This behaviour is characteristic of the solutions across the range of aspect ratios.

Through a modification of the optimum aspect ratio parameter $a$ in \cref{sigma_z_for_order_one}, there is a corresponding change in the aspect ratio felt in the discrete pressure matrix of elements in dry regions.
The parameter $a$ is varied over the range
$a\in[{10^{-4}},10^{4}]$ in a suite of 1001 simulations of the base Balzano case.
Solver iteration number is used as an indicator of conditioning, and plotted in
\cref{fig:balzano_a} for both the pressure and velocity calculations as mean and maximum values over the course of a WD phase.
The parameter range has been spaced equally in log-space in order to give a good representation of the behaviour over the large range of domain aspect ratios.
This is achieved with a discrete parameter space defined for a parameter $p$, such that
\begin{equation}
p \in
\{
10^{%
\left(
\left(
2s / (n - 1)
- 1
\right)
m
\right)
}
\!
:
s \in \mathbb{Z}, 0 \le s < n
\},
\nonumber
\end{equation}
for $n$, the number of distinct individual simulations spanning the parameter space over
$m$ orders of magnitude either side of zero,
such that
$p \in [10^{-m},10^{m}]$.

Whilst the conditioning of the velocity solver is largely unaffected, the number of iterations required for pressure convergence increases dramatically as the magnitude of the parameter $a$ increases.
As the aspect ratio parameter becomes acutely large with $\abs{a}\to\infty$, behaviour tends to that of the system without the scheme applied.
It is clear that the vertical relaxation scheme has a positive impact on conditioning, reducing the number of required iterations in the pressure solution in this test by a factor of 20.
With Picard iteration numbers also reduced as a consequence, this effect is multiplied for significant overall performance gains.

Changes in the parameter demonstrate that the scheme significantly improves conditioning in the base case.
Now an optimal aspect ratio $a\!=\!1$ is specified and actual changes to the domain extents considered.
Again a suite of simulations are run to span the parameter space and determine conditioning,
and the number of iterations required for convergence of the pressure is shown in
\cref{fig:balzano_its}, with and without the relaxation conditioning.
In the range considered, the improvement is reduced by a factor of up to 20 and results highlight that the approach eliminates a dependence of conditioning on aspect ratio.

\subsection{Second Balzano shelf channel benchmark}
\label{dztests}
\begin{SCfigure*}
\centering
\includegraphics[width=13.9cm]{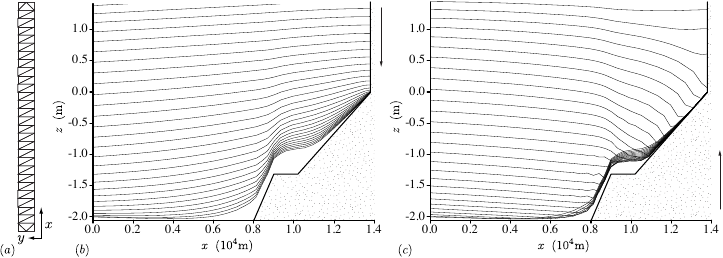}
\vspace{-0.3cm}
\caption{%
The second Balzano channel flow benchmark with a horizontal extent, in this presented case,
of $1.38\times{10}^4\textrm{m}$,
corresponding to a minimum element aspect ratio of ${\sim{10}^{-6}}$.
The discretised horizontal surface (a)
contains 58 triangular elements
with a characteristic length scale ${\sim50-100\textrm{m}}$.
Vertical sections show the free surface position at 10min intervals
for the initial drying phase (b) and during wetting (c).
}
\label{fig:balzano2}
\end{SCfigure*}
This case also originates in Balzano~\cite{balzano98}
and differs from the first by the inclusion of a shelf break in the sloped bathymetry,
defined in \autoref{sec:secondbalzano}.
The horizontal domain is discretised
in a way to ensure accurate bottom boundary representation,
such that element faces align with the discontinuous changes in surface gradient
(\cref{fig:balzano2}).
Except for the change in bathymetry, discretisation proceeds in the same manner as the first Balzano case of \cref{sec:balzano1},
and is again run over a range of aspect ratios.

The free surface evolution in the case with minimum element aspect ratio ${10}^{-6}$
is
shown in \cref{fig:balzano2}, again characteristic of the formulation over the range of aspect ratios.
In addition to the oscillatory and retention problems already mentioned, Balzano noticed a runoff problem with some methods in this test case,
where water remains on the shelf during the dry period instead of flowing into the basin.
Like
\cite{funke11} and \cite{gourgue09}, the runoff is observed to be linear in time, the correct physical behaviour.

With the irregular bathymetry of this case, we consider the effect of how the length scales that are passed to the relaxation scheme are calculated, as discussed in
\cref{sec:lengthscales}.
The characteristic height $\Delta z$ varies both by element and over elements, and can be calculated at quadrature points for increased accuracy.
Noting the role of these length scales in the vertical velocity relaxation inverse time scale
\cref{sigma_z_for_order_one},
we see that errors in how they are determined
influence conditioning in the same manner as that of perturbations of $a$
from the optimum value of 1,
except to a greater degree due to the quadratic dependence
which, following
\cref{fig:balzano_a},
reduces the effectiveness of the conditioning.

\begin{table}[h]
\centering
\begin{tabular}{lrrrr}
\toprule
Method & \multicolumn{2}{c}{Drying phase} & \multicolumn{2}{c}{Wetting phase} \\
\cmidrule(r){2-3} \cmidrule(r){4-5}
 & max & mean & max & mean \\
\midrule
Minimum & 512 & 475 & 512 & 475 \\
Maximum & 305 & 303 & 281 & 269 \\
Mean & 287 & 285 & 328 & 310 \\
Minimum capped & 300 & 297 & 321 & 301 \\
Jacobian & 310 & 281 & 264 & 259 \\
\bottomrule
\end{tabular}
\vspace{-0.1cm}
\caption{%
Pressure solver iterations to convergence in the
second Balzano benchmark with a domain aspect ratio of
$3.62\times{10}^{-6}$
for five approaches to calculating characteristic height.
}
\label{tab:dzcalc}
\end{table}
Five approaches are considered
(\cref{tab:dzcalc,sec:lengthscales}).
The methods `minimum', `maximum' and `mean' each refer to the
minimum, maximum and mean of the set of six vertical lengths $\Delta z$ calculated from the four tetrahedral element vertices.
The minimum of these performs poorly in all phases, so its value was limited by a lower bound in the `minimum capped' approach, which prevents the applied absorption becoming too large.
This produced better conditioning than the maximum in the drying phase,
and whilst improved in the wetting phase, the maximum here still produced better conditioning.
The mean behaves very well in the drying phase, but only satisfactorily during wetting.
This implies all three of these norms are not capturing all of the important parameters to determine an optimum $\Delta z$.
The Jacobian approach using the determinant of a contracted transformation matrix at quadrature points provides the best conditioning during the wetting phase.
The conditioning in the drying phase is not consistently the best, but the lowest mean number implies it is best overall.
In the Balzano shelf case examined here, the number of iterations required for convergence is approximately halved by a careful consideration of the calculation of $\Delta z$.

\subsection{Thacker parabolic basin benchmark}
\label{sec:thacker}
The Thacker parabolic bowl \citep{thacker81} is
an idealised ocean basin that thins at its edges,
with bathymetry defined in \cref{sec:thackerdescription}.
It is a challenging free surface flow problem with WD
that has previously been used in intercomparison studies \cite{balzano98,funke11,gourgue09,karna11}.
An analytical solution for the evolution of the free surface is known
(also in \cref{sec:thackerdescription})
when both dissipation and Coriolis are absent,
and the case suitable for the evaluation of
spatial and temporal accuracy, and volume conservation.
\begin{figure}[!ht]
\centering
\includegraphics[width=8.0cm]{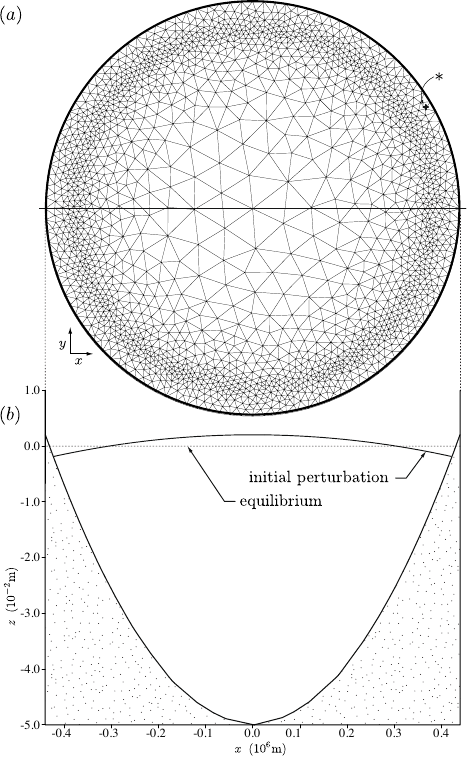}
\vspace{-0.3cm}
\caption{%
Thacker parabolic basin benchmark with a vertical extent of 5cm.
The discretised horizontal surface (a)
follows the metric \cref{thackermetric}
for the case
$\Delta x \!=\!{10}^4\textrm{m}$,
containing 1,818 nodes and 3,750 triangular elements
and characteristic length scales
${\sim\!10-100\textrm{km}}$.
Along the bisecting line,
(b) shows the
representation of the parabolic bathymetry
together with equilibrium and initial perturbed free surface positions.
}
\label{fig:thacker}
\end{figure}
\begin{figure}[!ht]
\centering
\includegraphics[width=7.5cm]{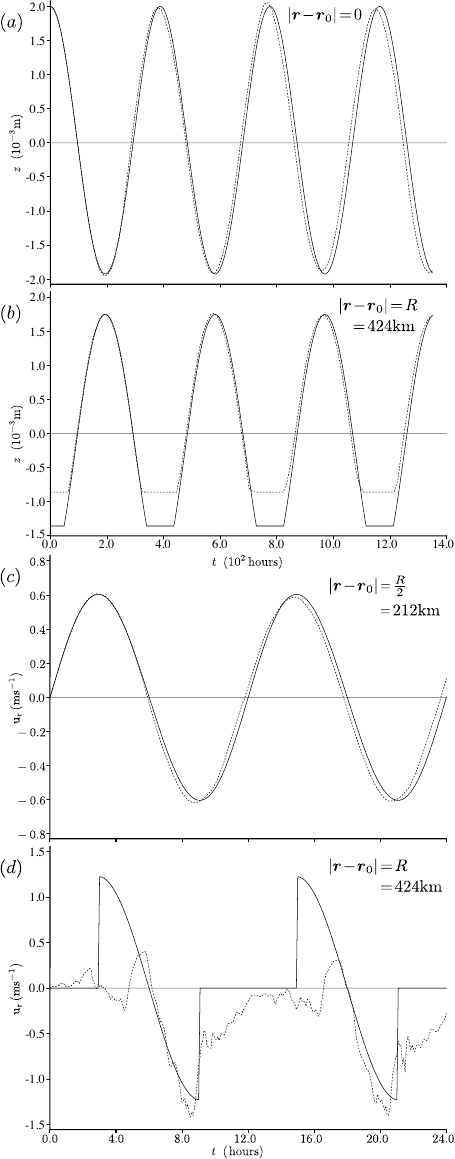}
\vspace{-0.3cm}
\caption{%
Thacker parabolic benchmark, showing analytical (solid) and numerical (dashed) solutions.
Evolution of $\eta$ in the
$5.68\times{10}^{-8}$
aspect ratio domain
at
(a) the centre of the basin
and (b) a distance 424km from the centre
marked * in \cref{fig:thacker}(b).
Radial velocity $u_\mathrm{r}$ evolution at the free free surface,
in the base domain
of \cref{fig:thacker},
at a distance
(c) 212km, and (d) 424km.
}
\label{fig:thackerevolution}
\end{figure}
\begin{figure}[!ht]
\centering
\includegraphics[width=6.5cm]{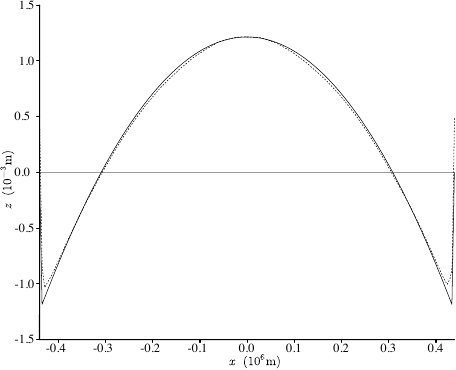}
\vspace{-0.3cm}
\caption{%
Thacker benchmark analytical (solid) and numerical (dashed) $\eta$ solutions
along the vertical slice indicated in \cref{fig:thacker}
after thirty days, in a domain with global aspect ratio
$5.68\times{10}^{-8}$.
}
\label{fig:thackerfreesurface720h}
\end{figure}
\begin{SCfigure*}
\centering
\includegraphics[width=11cm]{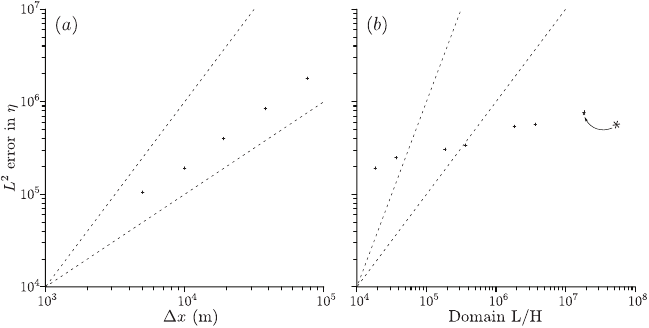}
\vspace{-0.3cm}
\caption{%
Thacker benchmark convergence properties.
Accuracy relative to (a) edge length in base domain of \cref{fig:thacker}
and with respect to aspect ratio in (b).
Error is evaluated at the point of time that the initial wetting period is complete.
Linear and quadratic gradients are indicated by dashed lines.
The diagonal cross, marked by *, points to a case with a relative tolerance reduced to $10^{-10}$.
}
\label{fig:thackerconvergence}
\end{SCfigure*}

The base case domain size matches that of \cite{thacker81,balzano98,gourgue09,funke11,karna11} with
a
$880\textrm{km}$ horizontal extent,
$R\!=\!430.62\textrm{km}$,
$h_0\!=\!50\textrm{m}$,
$\eta_0\!=\!2\textrm{m}$,
with a minimum water thickness of
$d_0\!=\!0.5\textrm{m}$.
No viscosity or drag terms result in a non-damped free surface oscillation with a 12h period.
We make the assumption that in this domain
the hydrostatic component of the free surface perturbation dominates with the non-hydrostatic part small, and thus the solution converges to the analytical function in \cref{sec:thackerdescription}.

Conditioning is examined for domain aspect ratios ranging over four orders of magnitude,
from the base
$5.68\times{10}^{-5}$
down to
$5.68\times{10}^{-8}$.
This is
achieved through vertical scaling the domain and $d_0$,
with the maximum equilibrium water column depth varying
between $50\textrm{m} - 5\textrm{cm}$.
With the characteristic horizontal edge length
${\sim{10}^{4}\textrm{m}}$ close to the edges where the domain dries,
element aspect ratios vary similarly
${\sim5\times{10}^{-5}} - 5\times{10}^{-8}$.
A cross section of the resulting single-layer basin domain
for the
$h_0\!=\!5\textrm{cm}$ case
is shown in
\cref{fig:thacker}(b), with the initial perturbation
$\eta_0$
ensuring a minimum thickness of $d_0$ is applied.

Edge element length scales are defined isotropically by
\begin{equation}
\varepsilon(\boldsymbol{r}) =
\Delta x
\left(
9
\abs{
{(R - \abs{\boldsymbol{r} - \boldsymbol{r}_0})}/{R}
}
+ 1
\right)
,
\label{thackermetric}
\end{equation}
which
for the case
$\Delta x\!=\!{10}^4\textrm{m}$ in \cref{fig:thacker}, result in a
range from 100km in the middle
down to 10km at a distance $3.8\times{10}^5\textrm{m}$ from the centre,
in an approach following \cite{funke11}.

Numerical evolution of the free surface for the highest aspect ratio case, shown in \cref{fig:thackerevolution}(a)-(b),
is observed to fit the analytical solution very well, even with elements of a very high aspect ratio ($5\times{10}^{-8}$).
Like the results from the more modest domain size a phase shift is observed, which also seen in \cite{funke11}, is a feature produced by the thin layer in the dry areas.
We can eliminate numerical dissipation inherent in the scheme as a contributor,
as we find that with solves iterated to convergence, volume is conserved up to a relative factor of
$1.0\times{10}^{-11}$, which is attributed to numerical round off error.
This phase shift is reduced with an increase in mesh resolution \citep[see][]{funke11}, which contributes to the increase in accuracy observed in
\cref{fig:thackerconvergence}(b).
In the time series taken at the edge of the domain,
it is clear when the location becomes dry in both the analytical and numerical solution,
and where the factor of $d_0$ is maintained in the latter (here 0.5mm).

The radial velocity at the free surface at two locations is presented
in \cref{fig:thackerevolution}(c)-(d)
at approximately the same relative locations as those considered in \cite{casulli07}, and is compared to the
analytical solution provided in
\cref{sec:thackerdescription}.
In the main body of fluid the solution is a very good match, with the same shift
observed in $\eta$ as in \cref{fig:thackerfreesurface720h} above and \cite{funke11}.
Close to the edge of the basin, $\boldsymbol{u}_r$ is not as well predicted as $\eta$.
This is partly due to the continuous nature of the thin-film approach, which solves for
$\boldsymbol{u}$
in both wet and dry regions.
The spatial discretisation local to this point is relatively coarse, and additionally is not aligned to a radial direction, which makes
$\boldsymbol{u}_r$ particularly challenging to calculate.
This and the phase error, can be mitigated by increasing spatial resolution and constraining mesh structure to align with flow direction in inundation regions.
Importantly, accuracy of \cite{funke11} is maintained, whilst the difficulty in solving the linear systems is much reduced.

The position of the free surface
in a vertical slice of the domain along the line indicated in \cref{fig:thacker}
and
after a period of
thirty days, to include two each of the WD phases, is shown in \cref{fig:thackerfreesurface720h}.
Spatially, the numerical solution is a good fit to the analytical solution and its resolution of the WD front comparable to studies in more modest aspect ratio domains
\cite{gourgue09,funke11}.
The use of the vertical velocity relaxation approach and iterative solvers for the linear systems does not have a significant impact on accuracy of the solution,
and provides a formulation for high aspect ratio domains that performs as well as those of modest size.

An evaluation of error
$e(t)$,
at a time $t$,
is made under an
$L_2$ norm of the absolute difference,
such that
\begin{align}
e(t)
&=
\norm{
\eta(t)
-
\mathrm{max}
\left(
\eta_a(t), h + d_0
\right)
}_{2,\Omega}%
,
\nonumber
\\[-8pt]
&=
\left(
\int_\Omega
\abs{
\eta(t)
-
\mathrm{max}
\left(
\eta_a(t), h + d_0
\right)
}^2
\right)^{\frac{1}{2}}\!\!,
\nonumber
\end{align}
for
$h$ and $\eta_a$ defined in
\cref{sec:thackerdescription}.
The minimum water depth is included in the analytical solution, since this
is the free surface height the formulation converges to,
and the domain $\Omega$ encompasses both wet and dry regions.

Solution convergence with respect to the smallest horizontal characteristic edge length $\Delta x$ is considered in \cref{fig:thackerconvergence}(a) for the base domain,
where the time step is linearly scaled to maintain a constant CFL number.
Meshed domains are generated by scaling the metric~\cref{thackermetric}.
With this WD formulation we obtain the linear convergence in error to characteristic edge length observed in \cite{funke11}.

The impact of domain aspect ratio on the accuracy of the calculation of free surface height after the initial wetting phase
is considered in
\cref{fig:thackerconvergence}(b).
Notably the error does not increase significantly with an increase in the magnitude of the aspect ratio and is far from linear.
The increase can be accounted for, to some extent, by the fixed relative tolerance on the iterative solvers of the linear systems.
Adjusting this tolerance to increase convergence in cases with very small edge lengths could help to increase accuracy at this level.
A small improvement in accuracy is seen in the highest aspect ratio case considered in
\cref{fig:thackerconvergence}(b)
where the relative error tolerance of ${10}^{-7}$ described in \cref{sec:application} is reduced to ${10}^{-10}$.
It is a significant result that a solution can be found for these cases with very high aspect ratios
and additionally, that the approach does not have an appreciable impact on accuracy.

\subsection{Basin inundation}
\begin{figure}[!ht]
\vspace{-0.2cm}
\centering
\includegraphics[width=8.3cm]{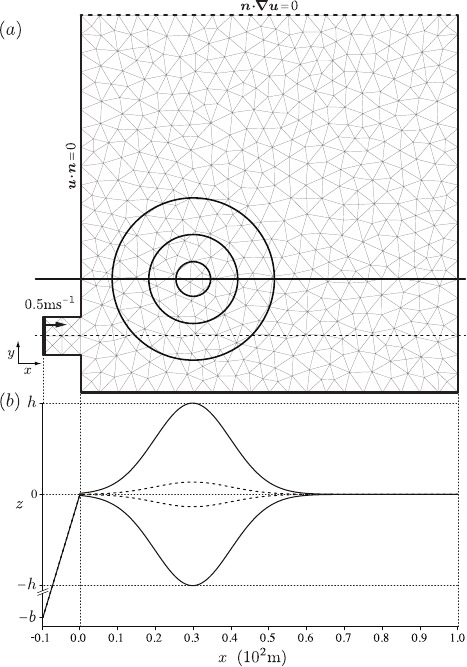}
\vspace{-0.3cm}
\caption{%
Flood plain basin benchmark (a) horizontal domain.
Circular contours mark
$\frac{1}{10}h_0$,
$\frac{1}{2}h_0$ and
$\frac{9}{10}h_0$, of the applied bathymetric features.
Solid and dashed lines mark the bathymetry cross-sections appearing in (b),
where plain, hollow and hill case profiles are shown.
}
\label{fig:basinmesh}
\end{figure}
This case considers the inundation of water into an initially dry basin,
with the effect of bathymetric features on WD front propagation also examined.
The base domain is shown in
\cref{fig:basinmesh} and
consists of a basin
with horizontal extent $100\textrm{m}\times100\textrm{m}$,
and an inlet of width 10m, its centre positioned 15m in from one of the corners.
The domain is discretised with elements of a characteristic edge length of 5m.
The problem is forced with a normal inlet velocity of $0.5\textrm{ms}^{-1}$ to model a levee breach into a flood plain on an urban scale.

To provide a more natural forcing, instead of applying a flux directly on the boundary, the inlet is extended back 10m and is
maintained wet throughout by developing a sloped bathymetry back, down to a depth of $b\!=\!20\textrm{m}$, as seen \cref{fig:basinmesh}(b).
The normal inlet velocity is then applied to the face that has been extended back, with velocity slip conditions on the adjacent sides.
This was found to avoid problems with the inflow at the edges of the breach.
At the outflow on the far boundary at $y\!=\!100\textrm{m}$,
a natural Neumann condition is applied perpendicular to the boundary,
such that
${\partial v}/{\partial y} = 0$, for velocity $v$ in the $y-$direction.
All other boundaries are closed, with no normal flow conditions applied.
Other velocity components are free and left unconstrained.

To ensure accuracy of the calculation of prognostic variables is not affected by the use of relatively large time steps with potential impact on the conditioning analysis,
\ts is set conservatively at 10s to give a maximum Courant number of 1.

\begin{figure}[!ht]
\centering
\includegraphics[width=7.9cm]{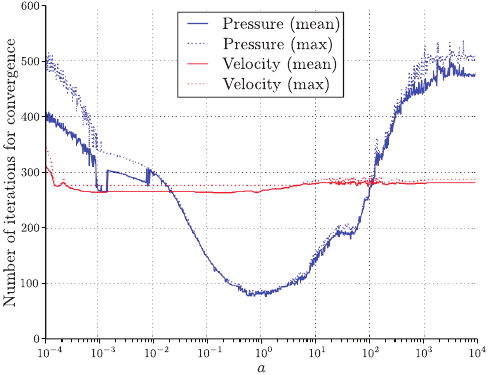}
\vspace{-0.3cm}
\caption{%
Impact of the optimal aspect ratio parameter on solver conditioning in the flood plain basin benchmark
over a WD phase. 1001 individual simulations shown.
}
\label{fig:basin_a}
\end{figure}
\begin{figure}[!ht]
\centering
\includegraphics[width=7.9cm]{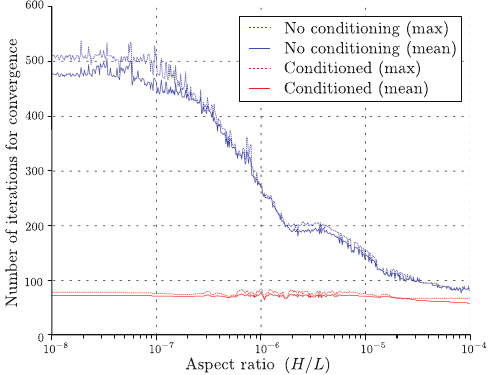}
\vspace{-0.3cm}
\caption{%
Pressure solver iterations to convergence in the flood plain basin benchmark with respect to domain global aspect ratio,
in 1001 individual simulations,
with and without conditioning applied.
}
\label{fig:basin_its}
\end{figure}
\begin{figure}[!ht]
\centering
\includegraphics[width=7.9cm]{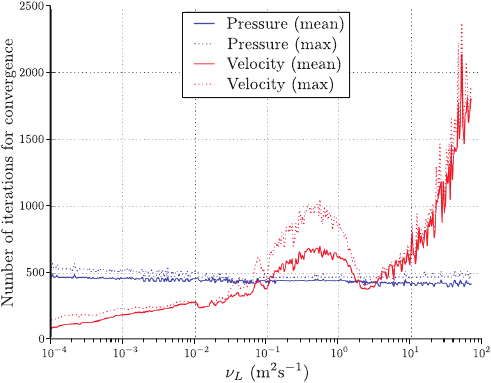}
\vspace{-0.3cm}
\caption{%
Impact of horizontal viscosity $\nu_L$ of \cref{big_v}
on solver conditioning in the flood plain basin with hill protrusion benchmark over a WD phase.
1001 individual simulations shown.
}
\label{fig:basin_vis}
\end{figure}

\begin{figure*}
\centering
\includegraphics[width=17cm]{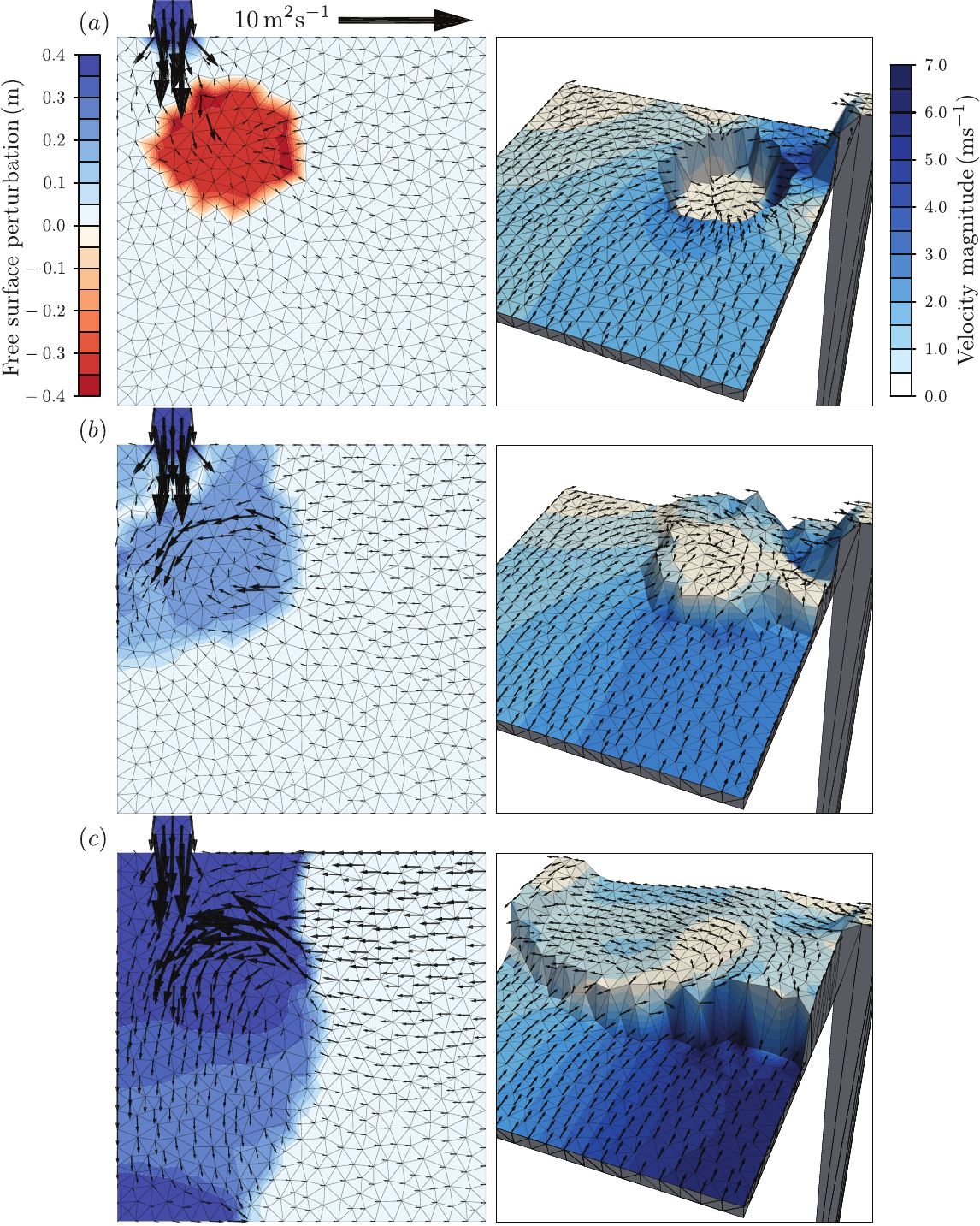}
\vspace{-0.3cm}
\caption{
Inundation into a flood basin of side length
$100\textrm{m}$ and threshold value $1\textrm{cm}$
containing a hollow bathymetric feature.
Three successive visualisations with (a) the hollow filling, (b) a hydraulic jump and (c) propagation further into the plain, are shown
at 7350, 20400 and 37650s into the simulation, respectively.
The left panels contain contour plots of free surface perturbation, overlaid with magnitude-scaled vectors of depth-integrated velocity.
The right panels present the 3D domain stretched in the vertical by a factor of $40$, to better show the change in free surface height,
with the inlet breach and connecting reservoir seen on the right side.
Contours of the magnitude of surface velocity are plotted together with vectors indicating the surface flow direction.
Note the velocity fields presented are those in a continuous \po space,
calculated through a Galerkin projection from the discontinuous \podg prognostic velocity field \cite[see][]{candy08}.
}
\label{fig:dike}
\end{figure*}

In a similar approach taken for the Balzano slope case, we consider the influence of the
optimum aspect ratio parameter $a$ on conditioning in the base domain with aspect ratio ${10}^4$,
over a parameter space spanned by 1001 simulations shown in
\cref{fig:basin_a}.
Conditioning of the pressure solver is significantly improved, by over a factor of six in this modest aspect ratio case.
Again velocity is only slightly affected, and felt through the coupling, a consequence of better pressure conditioning.
When varying simulation domain extent, with an optimal choice of $a\!=\!1$, similar behaviour is observed and shown in \cref{fig:basin_its}.

In practice large gradients in bathymetry have an impact on conditioning.
This is studied with the introduction of a depression in the domain to form a hollow and conversely, a raised hill.
Both interact differently with the incoming wetting front.
These features are introduced to the domain with a
Gaussian perturbation, which is defined at all points $\boldsymbol{r}$ on the horizontal surface of the domain by
$
h(\boldsymbol{r})
=
h_0
e^{
-\frac{1}{2}
\left(
(\boldsymbol{r} -
\boldsymbol{r}_0)
\bcdot
\bar{\boldsymbol{\sigma}}
\right)^2
}\!\!,
$
for $h_0$ the maximum deviation in height, which occurs at the centre where
$\boldsymbol{r}\!=\!\boldsymbol{r}_0$.
The inverse variance vector $\bar{\boldsymbol{\sigma}}$ defines width, and consequently the gradient, of the obstacle.
In the scales of the base case, the magnitude of the perturbation $\abs{h_0}$ is 5m,
with a width of 10m, defined by
$
\bar{\boldsymbol{\sigma}}\!=\!\left(
\frac{1}{10}, \frac{1}{10}
\right)
$.
The perturbation is positioned at 30m in from each of the bounding edges at the corner closest to the inlet.
At the start, the minimum water thickness of $d_0\!=\!1\textrm{cm}$ is applied above the bathymetry, to provide the initial thin dry flood basin.

The above is used to generate an inundation into a domain containing a large hollow with $h_0\!=\!-5\textrm{m}$.
Conditioning is further decreased with the presence of the hollow, with a mean number of 105 iterations required in pressure for the modest aspect ratio case.
Compared to the flat case, the number of iterations required increases at a greater rate,
and the positive effect on conditioning of the vertical relaxation scheme is further pronounced.
Additionally, the large gradients in bathymetry adversely affect conditioning of the velocity solver early in the simulation where large velocities develop around the steep slopes to fill the hollow.
This can be seen in the example snapshot results shown in \cref{fig:dike}.
Initially flow is strong from the breach, and predominantly flows into the hollow, whose surface oscillates in a similar manner to that seen in the Thacker parabolic bowl benchmark of \cref{sec:thacker}.
Once the hollow is filled, the free surface peaks and a hydraulic jump develops between the fast-flowing inlet from the breach and the formed lake.
The fluid then gains momentum in the direction of the inlet flow and is seen to build up on the opposite boundary.
A clear front has developed by this stage and begins to propagates across the plain towards the open boundary.
It is also possible to see the larger velocities that develop at the front and ahead in the thin dry regions.
This is motivation for the application of velocity conditioning discussed in the following.

In the case of the hill, with $h_0\!=\!5\textrm{m}$ in the base domain, the effect on velocity is more significant, particularly as the WD front meets the bathymetric intrusion.
In this case it is necessary to apply a regularisation to the momentum equation to improve conditioning, which is achieved through an application of a bulk volume viscosity, as introduced in \cref{sec:viscosityindry}.
The domain-wide horizontal viscosity $\nu_L$, is varied over the range
$\nu_L \in [{10}^{-4},{10}^{4}]\, \textrm{m}^{2}\textrm{s}^{-1}$
through \cref{big_v}
in the mid aspect ratio ${10}^{-6}$ case, with conditioning shown in
\cref{fig:basin_vis}.
As a general trend, the number of iterations required increases with strengthening of the horizontal viscosity.
There is however a point at which there is a noticeable dip, where the increase in intensity improves conditioning.
This decrease in the mean number of iterations is due to improvement of conditioning made when the front approaches and traverses the hill protrusion.
Limiting application of this conditioning to dry regions and its proximity, as described in
\cref{sec:viscosityindry},
allows WD fronts to encounter steep changes in bathymetry without the corresponding impact on conditioning of the velocity solver, in this implicit and continuous WD formulation.

Increasing the bottom drag through the Manning-Strickler parameterisation in this region as outlined in \cref{sec:big_drag} also acts to improve conditioning.
In particularly high aspect ratio cases with steep bathymetry, the velocity solver is too badly conditioned for efficient solution with a SSOR-GMRES iterative process without approaches such as the horizontal viscosity and drag
discussed.

\section{Conclusion}
\label{sec:conclusions}
In this paper a novel approach to efficiently modelling WD inundation processes
in \threed, capturing non-hydrostatic and baroclinic physics,
in the high aspect ratio domains that characterise geophysical systems has been proposed.

This has identified the ill-conditioning present in
implicit continuum WD methods applied in
fully \threed fluid flow models.
Following a quantification of the highly spatial and temporally variable contributing factors,
regularisation of the governing weak form
leads to a linear system that appears as
a unit aspect ratio problem.
The result is that the approach can be used to model WD in multiscale geophysical domains,
seamlessly alongside other challenging physics,
such as baroclinic and non-hydrostatic flow,
without a severe and limiting impact on
the iterative solvers typically required for efficient simulation of multi-physics \threed dynamics.

The approach has been demonstrated effective over a wide range of spatial scales
and correspondingly,
aspect ratios.
The predicted behaviour on convergence is verified in numerical tests
in both domain and element aspect ratios
representing up to 8 orders of magnitude difference,
with
discrete domains containing spatial scales
spanning 10 orders of magnitude.

The approach imposes no restrictions on space and time discretisation, permitting an arbitrarily flexible mesh choice (including generalised vertical coordinates), order of representation and implicit time integration.
All are important for system models simulating over a range of scales and physics.
Discretisation can be chosen largely independent of WD considerations,
with for example, spatial resolution focused on local physics modelling demands.

Use of a combined \fp variable strictly enforces consistency between the full \threed pressure and free surface perturbation.
Notably there is no need to interpolate $\eta$ and its derivatives from $\Gamma_\eta$ to the internal domain $\Omega$ for inclusion in the momentum calculation.
Consistency with other fields and conservation are achieved by the overall FE approach,
which can provide a high order continuum representation.
\popo and the heterogeneous element pairing \podgpt have been applied in the numerical tests.
The implicit treatment of \fp is inherited by $\overline{p}$ and $\eta$, and
as a result, \ts may be based solely on accuracy considerations and not stability when considering free surface wave propagation.
As discussed in \cite{dalpaos07} this may need careful consideration when
a system is under-resolved with a relatively irregular bottom topography containing sharp gradients,
or in high Froude number rapidly varying flow.

A limitation to note is that the free surface interface cannot become unduly complicated, including folds, since
the function $\eta$ is by definition injective
with only a single point permitted to lie on the surface for any point within the domain.
As such it is not possible to model breaking waves, a common limitation to all of the Eulerian approaches cited.

Unlike schemes applying additional viscosity or bed friction based on empirical numerical measures
that potentially lead to stabilisation through unphysical means,
the approach ensures physical consistency
such that
resultant
solutions are enforced to exist
in the space of solutions available to the original physically based weak form of the continuum governing equations \cref{momentumequation_nohydrostatic}--\cref{incompressibility2}.
Physical consistency is verified in the numerical tests.
Lastly, since the terms introduced specifically to improve conditioning are formulated in the continuum primitive form,
this part of the approach could equally be applied in other WD implementations for an arbitrary underlying discretisation.

This approach will not be optimum for some WD problems,
particularly due to the computational cost even with the aspect ratio problem solved,
where a single layer SWE approximation is sufficient, or computational efficiency may demand lower order methods for real-time tsunami prediction, for example.
However this approach now enables the modelling of physical phenomena not possible previously,
particularly those at the interfaces of traditionally separate fields.
With rapid ongoing development of computational resources,
this approach and similar will grow in use and become more common practice
--
a way to bring WD to seamless massive multiscale multi-physics Earth system models.

\section*{Acknowledgements}
\noindent
I would like to thank Christopher Pain for the helpful discussions at all stages of this research, as well as his reviews of the manuscript and his encouragement to publish.
Additionally I am grateful for Matthew Piggott's feedback and support for this work.
I am grateful to the three anonymous reviewers for their valuable comments and suggestions which helped improve the manuscript.
I would also like to acknowledge support from Fangxin Fang who helped secure the European Commission Framework Programme 7 PEARL grant (Ref 603663), which partly funded this work.
Further support came from
the UK Natural Environment Research Council (grant NE/G018391/1)
and
Engineering and Physical Sciences Research Council (grant EP/I00405X/1).
Computational resources were provided by
Imperial College
and
UK academic HECToR/ARCHER
HPC services.

\appendix
\renewcommand*{\thesection}{\Alph{section}}

\section{Kinematic condition and nomenclature}
\label{sec:kinematic}
\noindent
Evolution of the free surface
accommodating surface waves requires
a further prognostic variable
defining its height
$\eta\!: \Omega \mapsto \mathbb{R}$
(see \cref{fig:domain})
with the interface parametrised by
$z\!=\!\eta(x,y)$,
where
$\eta\!=\!0$
when the fluid is at rest and in equilibrium.
Without loss of generality, the reference frame is rotated to align $z$
to the local gravitational direction, with $\boldsymbol{n}_g\!=\!(0,0,1)^T$.
An additional constraint is required
and the
assumption made that a fluid parcel on the free surface remains there throughout time \citep{acheson90},
which with the coordinates of a fluid parcel
$(x(t), y(t), z(t))^T$,
is written
$\eta(x, y, z, t)\!=\!z, \; \textrm{for} \; t \in [0, T)$,
with time derivative
\begin{equation}
\frac{\partial \eta}{\partial t}
=
-
\frac{\partial \eta}{\partial x}
\frac{\partial x}{\partial t}
-
\frac{\partial \eta}{\partial y}
\frac{\partial y}{\partial t}
+
\frac{\partial z}{\partial t}
=
\boldsymbol{\overline{n}}
\bcdot \boldsymbol{u},
\nonumber
\end{equation}
for the surface normal vector
$
\boldsymbol{\overline{n}}
\!=\!
(
-{\partial \eta}/{\partial x},
-{\partial \eta}/{\partial y},
1
)^T
$
and normalised form
$\boldsymbol{n}\!=\!\boldsymbol{\overline{n}} / \left| \boldsymbol{\overline{n}} \right|$.
Scaling by
$\left| \boldsymbol{\overline{n}} \right|$ and noting
$\boldsymbol{\overline{n}} \bcdot \boldsymbol{n}_g\!=\!1$, gives
the kinematic condition
\begin{equation}
\boldsymbol{n} \bcdot \boldsymbol{n}_g
\frac{\partial \eta}{\partial t}
=
\boldsymbol{n} \bcdot \boldsymbol{u} \quad \textrm{on} \; \Gamma_\eta.
\nonumber
\end{equation}

\section{Finite element basis definitions}
\label{sec:spatial}
\noindent
The weak form \cite{brenner94} of the governing equations is obtained by
an inner product with all test basis functions from
a Sobolev space
$\mathcal{S}\!:= \mathcal{H}^1(\Omega)$
defined over the domain $\Omega$
with generalised first derivatives and an
$\mathcal{L}^2$ inner product.
The Galerkin FE spatially discretised equations are
found by limiting $\mathcal{S}$ to a discrete subspace
$\mathcal{S}^h\!\subset\!\mathcal{S}$, itself defined over a discrete representation of the domain,
containing a finite number of spanning orthogonal trial functions.
The prognostic variables are represented
\begin{equation*}
\boldsymbol{u}
\!:=
\sum u_i \boldsymbol{\phi}_i,
\;\;
p
\!:=
\sum p_j \psi_j,
\;\;
\textrm{for} \;
u_i, p_i \in \mathbb{R},
\end{equation*}
and trial functions
$\boldsymbol{\phi}_i\!: \Omega \mapsto \mathbb{R}^3$
and
$\psi_j\!: \Omega \mapsto \mathbb{R}$,
and sum over the entire Sobolev spaces.
Applications in this paper operate on meshes consisting of tetrahedral elements;
with
discontinuous piecewise linear functions
and
continuous piecewise quadratic functions for velocity and pressure respectively, referred to as
\podgpt
and introduced in
\cite{cotter09}.

\section{Second Balzano benchmark bathymetry}
\label{sec:secondbalzano}
\noindent
\begin{equation}
h(x) =
\left\lbrace
\begin{array}{ll}
x/2760, &
 \textrm{for} \;
 x \in [0.0, 3.6] \; \textrm{km},
\\
30/23, &
 \textrm{for} \;
 x \in (3.6,4.8] \; \textrm{km}, \;
\\
x/1380 - 50/23, &
 \textrm{for} \;
 x \in (4.8,6.0) \; \textrm{km}, \;
\\
x/2760, &
 \text{for} \;
 x \in [6.0,13.8] \; \textrm{km}. \;
\end{array}
\right.
\nonumber
\end{equation}

\section{Thacker parabolic basin benchmark functions}
\label{sec:thackerdescription}
\noindent
Basin bathymetry is a parabola of the form
\begin{equation*}
h(\boldsymbol{r}) = h_0
({R^2 - \abs{\boldsymbol{r} - \boldsymbol{r}_0}^2})/{R^2},
\end{equation*}
for position vector
$\boldsymbol{r}$ on the \twod horizontal surface,
$\boldsymbol{r}_0$ locating the disc centre,
$R$ the basin radius at rest, and
$h_0$ the equilibrium water column depth at
$\boldsymbol{r} = \boldsymbol{r}_0$.
The analytical free surface evolution, inferred from \cite{thacker81}, is
\begin{align}
&\eta_a(\boldsymbol{r}, t)
=
h_0
\!
\left(\!
\!
\frac{\sqrt{1\!-\!\hat{\eta}^2}}
{1 \!-\! \hat{\eta} \,\mathrm{cos}\, \omega t}
\!
-
\!
\frac{\abs{\boldsymbol{r} \!- \!\boldsymbol{r}_0}^2}{R^2}
\!\!
\left(\!
\frac{1-\hat{\eta}^2}
{(1 \!-\! \hat{\eta} \,\mathrm{cos}\, \omega t)^2}
\!-\!\!1\!\!
\right)
\!\!-\!\!1\!
\right),
\nonumber
\\
&
\qquad
\textrm{with}
\;\;\;
\hat{\eta}
=
\frac{(h_0 + \eta_0)^2 - h_0^2}
{(h_0 + \eta_0)^2 + h_0^2},
\;\;
\textrm{and}
\;\;
\omega^2
=
\frac{8 g h_0}{R^2},
\nonumber
\end{align}
where
$\hat{\eta}$
is the initial free surface perturbation at $\boldsymbol{r} = \boldsymbol{r}_0$,
such that
$\eta_a(\boldsymbol{r}_0, 0) = \hat{\eta}$.
Analytical horizontal velocities are calculated in \cite{thacker81}
(with polar versions in \cite{casulli07})
and for the examined cases reduce to
\begin{equation*}
u_r(\boldsymbol{r}) =
\frac{\omega \hat{\eta} \,\abs{\boldsymbol{r} \!- \!\boldsymbol{r}_0}  \,\mathrm{sin}\, \omega t}{2 (1 - \hat{\eta} \,\mathrm{cos}\, \omega t)},
\quad
 \textrm{for} \;
 \eta_a(\boldsymbol{r}, t) > 0.
\end{equation*}

\section*{Model availability}
\noindent
The approach is implemented in the general purpose, arbitrarily unstructured, FE
geophysics model
Fluidity \cite{piggott08},
\url{https://fluidity-project.org},
which is open source, available under LGPL at \url{https://github.com/FluidityProject/fluidity},
with verification tests specific to the approach for high aspect ratio domains described.

\begingroup
\raggedright
\bibliographystyle{model1-num-names}
\bibliography{wetdry}

\begin{thebibliography}{65}
\expandafter\ifx\csname natexlab\endcsname\relax\def\natexlab#1{#1}\fi
\providecommand{\url}[1]{\texttt{#1}}
\providecommand{\href}[2]{#2}
\providecommand{\path}[1]{#1}
\providecommand{\DOIprefix}{doi:}
\providecommand{\ArXivprefix}{arXiv:}
\providecommand{\URLprefix}{URL: }
\providecommand{\Pubmedprefix}{pmid:}
\providecommand{\doi}[1]{\href{http://dx.doi.org/#1}{\path{#1}}}
\providecommand{\Pubmed}[1]{\href{pmid:#1}{\path{#1}}}
\providecommand{\bibinfo}[2]{#2}
\ifx\xfnm\relax \def\xfnm[#1]{\unskip,\space#1}\fi
\bibitem[{Medeiros and Hagen(2013)}]{medeiros13}
\bibinfo{author}{S.~C. Medeiros}, \bibinfo{author}{S.~C. Hagen},
\newblock \bibinfo{title}{Review of {W}{D} algorithms for numerical tidal flow
  models},
\newblock \bibinfo{journal}{Int. J. Numer. Meth. Fl.} \bibinfo{volume}{71}
  (\bibinfo{year}{2013}) \bibinfo{pages}{473--487}.
\bibitem[{Oishi et~al.(2013)Oishi, Piggott, Maeda, Kramer, Collins, Tsushima,
  and Furumura}]{onishi13}
\bibinfo{author}{Y.~Oishi}, \bibinfo{author}{M.~D. Piggott},
  \bibinfo{author}{T.~Maeda}, \bibinfo{author}{S.~C. Kramer},
  \bibinfo{author}{G.~S. Collins}, \bibinfo{author}{H.~Tsushima},
  \bibinfo{author}{T.~Furumura},
\newblock \bibinfo{title}{Three-dimensional tsunami propagation simulations
  using an unstructured mesh {F}{E} model},
\newblock \bibinfo{journal}{J. Geophys. Res. Solid Earth} \bibinfo{volume}{118}
  (\bibinfo{year}{2013}) \bibinfo{pages}{2998--3018}.
\bibitem[{Cui et~al.(2014)Cui, Pietrzak, and Stelling}]{cui14}
\bibinfo{author}{H.~Cui}, \bibinfo{author}{J.~Pietrzak},
  \bibinfo{author}{G.~Stelling},
\newblock \bibinfo{title}{Optimal dispersion with minimized poisson equations
  for non-hydrostatic free surface flows},
\newblock \bibinfo{journal}{Ocean Modell.} \bibinfo{volume}{81}
  (\bibinfo{year}{2014}) \bibinfo{pages}{1--12}.
\bibitem[{Casulli and Stelling(1998)}]{casulli98}
\bibinfo{author}{V.~Casulli}, \bibinfo{author}{G.~S. Stelling},
\newblock \bibinfo{title}{Numerical simulation of 3{D} quasi-hydrostatic,
  free-surface flows},
\newblock \bibinfo{journal}{J. Hyd. Eng.} \bibinfo{volume}{124}
  (\bibinfo{year}{1998}) \bibinfo{pages}{678--686}.
\bibitem[{Casulli and Walters(2000)}]{casulli00}
\bibinfo{author}{V.~Casulli}, \bibinfo{author}{R.~A. Walters},
\newblock \bibinfo{title}{An unstructured grid, 3{D} model based on the shallow
  water equations},
\newblock \bibinfo{journal}{Int. J. Numer. Meth. Fl.} \bibinfo{volume}{32}
  (\bibinfo{year}{2000}) \bibinfo{pages}{331--348}.
\bibitem[{Defina(2000)}]{defina00}
\bibinfo{author}{A.~Defina},
\newblock \bibinfo{title}{2{D} shallow flow equations for partially dry areas},
\newblock \bibinfo{journal}{Water Resour. Res.} \bibinfo{volume}{36}
  (\bibinfo{year}{2000}) \bibinfo{pages}{3251--3264}.
\bibitem[{D'{A}lpaos and Defina(2007)}]{dalpaos07}
\bibinfo{author}{L.~D'{A}lpaos}, \bibinfo{author}{A.~Defina},
\newblock \bibinfo{title}{Mathematical modeling of tidal hydrodynamics in
  shallow lagoons: a review of open issues and applications to the venice
  lagoon},
\newblock \bibinfo{journal}{Comp. Geosci.} \bibinfo{volume}{33}
  (\bibinfo{year}{2007}) \bibinfo{pages}{476--496}.
\bibitem[{Lin et~al.(2004)Lin, Cheng, Edris, and Yeh}]{lin04}
\bibinfo{author}{H.-C.~J. Lin}, \bibinfo{author}{H.-P. Cheng},
  \bibinfo{author}{E.~V. Edris}, \bibinfo{author}{G.-T. Yeh},
\newblock \bibinfo{title}{Modeling surface and subsurface hydrologic
  interactions in a south {F}lorida watershed near the {B}iscayne {B}ay},
\newblock in: \bibinfo{editor}{C.~T. Miller}, \bibinfo{editor}{G.~F. Pinder}
  (Eds.), \bibinfo{booktitle}{15$^{\textrm{th}}$ Int. Conf. Comp. Meth. Water
  Resour.}, \bibinfo{year}{2004}, pp. \bibinfo{pages}{1607--1618}.
\bibitem[{Bates and Anderson(1993)}]{bates93}
\bibinfo{author}{P.~D. Bates}, \bibinfo{author}{M.~G. Anderson},
\newblock \bibinfo{title}{A two-dimensional finite-element model for river flow
  inundation},
\newblock \bibinfo{journal}{Proc. R. Soc. Lond. A} \bibinfo{volume}{440}
  (\bibinfo{year}{1993}) \bibinfo{pages}{481--491}.
\bibitem[{Chen et~al.(2003)Chen, Liu, and Beardsley}]{chen03}
\bibinfo{author}{C.~Chen}, \bibinfo{author}{H.~Liu}, \bibinfo{author}{R.~C.
  Beardsley},
\newblock \bibinfo{title}{An unstructured grid, {F}{V}, 3{D}, primitive
  equations ocean model: {A}pplication to coastal ocean and estuaries},
\newblock \bibinfo{journal}{J. Atmos. Ocean. Tech.} \bibinfo{volume}{20}
  (\bibinfo{year}{2003}) \bibinfo{pages}{159--186}.
\bibitem[{Oey(2005)}]{oey05}
\bibinfo{author}{L.-Y. Oey},
\newblock \bibinfo{title}{A {W}{D} scheme for {P}{O}{M}},
\newblock \bibinfo{journal}{Ocean Modell.} \bibinfo{volume}{9}
  (\bibinfo{year}{2005}) \bibinfo{pages}{133--150}.
\bibitem[{Begnudelli and Sanders(2006)}]{begnudelli06}
\bibinfo{author}{L.~Begnudelli}, \bibinfo{author}{B.~F. Sanders},
\newblock \bibinfo{title}{Unstructured grid finite-volume algorithm for
  shallow-water flow and scalar transport with {W}{D}},
\newblock \bibinfo{journal}{J. Hyd. Eng.} \bibinfo{volume}{132}
  (\bibinfo{year}{2006}) \bibinfo{pages}{371--384}.
\bibitem[{Bradford and Sanders(2002)}]{bradford02}
\bibinfo{author}{S.~F. Bradford}, \bibinfo{author}{B.~F. Sanders},
\newblock \bibinfo{title}{Finite-volume model for shallow-water flooding of
  arbitrary topography},
\newblock \bibinfo{journal}{J. Hyd. Eng.} \bibinfo{volume}{128}
  (\bibinfo{year}{2002}) \bibinfo{pages}{289--298}.
\bibitem[{Lynett et~al.(2002)Lynett, Wu, and Liu}]{lynett02}
\bibinfo{author}{P.~J. Lynett}, \bibinfo{author}{T.-R. Wu},
  \bibinfo{author}{P.~L.-F. Liu},
\newblock \bibinfo{title}{Modeling wave runup with depth-integrated equations},
\newblock \bibinfo{journal}{Coast. Eng.} \bibinfo{volume}{46}
  (\bibinfo{year}{2002}) \bibinfo{pages}{89--107}.
\bibitem[{Heniche et~al.(2000)Heniche, Secretan, Boudreau, and
  Leclerc}]{heniche00}
\bibinfo{author}{M.~Heniche}, \bibinfo{author}{Y.~Secretan},
  \bibinfo{author}{P.~Boudreau}, \bibinfo{author}{M.~Leclerc},
\newblock \bibinfo{title}{A 2{D} {F}{E} drying-wetting shallow water model for
  rivers and estuaries},
\newblock \bibinfo{journal}{Adv. Water Resour.} \bibinfo{volume}{23}
  (\bibinfo{year}{2000}) \bibinfo{pages}{359--372}.
\bibitem[{Jiang and Wai(2005)}]{jiang05}
\bibinfo{author}{Y.~Jiang}, \bibinfo{author}{O.~W. Wai},
\newblock \bibinfo{title}{Drying-wetting approach for 3{D} {F}{E} sigma
  coordinate model for estuaries with large tidal flats},
\newblock \bibinfo{journal}{Adv. Water Resour.} \bibinfo{volume}{28}
  (\bibinfo{year}{2005}) \bibinfo{pages}{779 -- 792}.
\bibitem[{Warner et~al.(2013)Warner, Defne, Haas, and Arango}]{warner13}
\bibinfo{author}{J.~C. Warner}, \bibinfo{author}{Z.~Defne},
  \bibinfo{author}{K.~Haas}, \bibinfo{author}{H.~G. Arango},
\newblock \bibinfo{title}{A {W}{D} scheme for {ROMS}},
\newblock \bibinfo{journal}{Comp. Geosci.} \bibinfo{volume}{58}
  (\bibinfo{year}{2013}) \bibinfo{pages}{54--61}.
\bibitem[{vant Hof and Vollebregt(2005)}]{vant05}
\bibinfo{author}{B.~vant Hof}, \bibinfo{author}{E.~A.~H. Vollebregt},
\newblock \bibinfo{title}{Modelling of {W}{D} of shallow water using artificial
  porosity},
\newblock \bibinfo{journal}{Int. J. Numer. Meth. Fl.} \bibinfo{volume}{48}
  (\bibinfo{year}{2005}) \bibinfo{pages}{1199--1217}.
\bibitem[{Ip et~al.(1998)Ip, Lynch, and Friedrichs}]{ip98}
\bibinfo{author}{J.~Ip}, \bibinfo{author}{D.~Lynch},
  \bibinfo{author}{C.~Friedrichs},
\newblock \bibinfo{title}{Simulation of estuarine flooding and dewatering with
  application to {G}reat {B}ay, {N}ew {H}ampshire},
\newblock \bibinfo{journal}{Est. Coast. Shelf Sci.} \bibinfo{volume}{47}
  (\bibinfo{year}{1998}) \bibinfo{pages}{119 -- 141}.
\bibitem[{K{\"a}rn{\"a} et~al.(2011)K{\"a}rn{\"a}, de~Brye, Gourgue,
  Lambrechts, Comblen, Legat, and Deleersnijder}]{karna11}
\bibinfo{author}{T.~K{\"a}rn{\"a}}, \bibinfo{author}{B.~de~Brye},
  \bibinfo{author}{O.~Gourgue}, \bibinfo{author}{J.~Lambrechts},
  \bibinfo{author}{R.~Comblen}, \bibinfo{author}{V.~Legat},
  \bibinfo{author}{E.~Deleersnijder},
\newblock \bibinfo{title}{A fully implicit {W}{D} method for {D}{G}-{F}{E}{M}
  shallow water models, with an application to the {S}cheldt {E}stuary},
\newblock \bibinfo{journal}{Comp. Meth. Appl. Mech. Engrg.}
  \bibinfo{volume}{200} (\bibinfo{year}{2011}) \bibinfo{pages}{509 -- 524}.
\bibitem[{Leendertse(1970)}]{leendertse70}
\bibinfo{author}{J.~J. Leendertse},
\newblock \bibinfo{title}{A water quality simulation model for well-mixed
  estuaries and coastal seas: volume {I}, principles of computation}
  (\bibinfo{year}{1970}).
\bibitem[{Balzano(1998)}]{balzano98}
\bibinfo{author}{A.~Balzano},
\newblock \bibinfo{title}{Evaluation of methods for numerical simulation of
  {W}{D} in shallow water flow models},
\newblock \bibinfo{journal}{Coastal Eng.} \bibinfo{volume}{34}
  (\bibinfo{year}{1998}) \bibinfo{pages}{83--107}.
\bibitem[{Stelling and Duinmeijer(2003)}]{stelling03a}
\bibinfo{author}{G.~S. Stelling}, \bibinfo{author}{S.~P.~A. Duinmeijer},
\newblock \bibinfo{title}{A staggered conservative scheme for every {F}roude
  number in rapidly varied shallow water flows},
\newblock \bibinfo{journal}{Int. J. Numer. Meth. Fl.} \bibinfo{volume}{43}
  (\bibinfo{year}{2003}) \bibinfo{pages}{1329--1354}.
\bibitem[{Walters(2005)}]{walters05}
\bibinfo{author}{R.~A. Walters},
\newblock \bibinfo{title}{Coastal ocean models: two useful {F}{E} methods},
\newblock \bibinfo{journal}{Contin. Shelf Res.} \bibinfo{volume}{25}
  (\bibinfo{year}{2005}) \bibinfo{pages}{775--793}.
\bibitem[{Stelling and Zijlema(2003)}]{stelling03b}
\bibinfo{author}{G.~Stelling}, \bibinfo{author}{M.~Zijlema},
\newblock \bibinfo{title}{An accurate and efficient finite-difference algorithm
  for non-hydrostatic free-surface flow with application to wave propagation},
\newblock \bibinfo{journal}{Int. J. Numer. Meth. Fl.} \bibinfo{volume}{43}
  (\bibinfo{year}{2003}) \bibinfo{pages}{1--23}.
\bibitem[{Casulli and Stelling(2010)}]{casulli10}
\bibinfo{author}{V.~Casulli}, \bibinfo{author}{G.~S. Stelling},
\newblock \bibinfo{title}{Semi-implicit subgrid modelling of 3{D} free-surface
  flows},
\newblock \bibinfo{journal}{Int. J. Numer. Meth. Fl.} \bibinfo{volume}{67}
  (\bibinfo{year}{2010}) \bibinfo{pages}{441--449}.
\bibitem[{Volp et~al.(2016)Volp, van Prooijen, Pietrzak, and Stelling}]{volp16}
\bibinfo{author}{N.~Volp}, \bibinfo{author}{B.~van Prooijen},
  \bibinfo{author}{J.~Pietrzak}, \bibinfo{author}{G.~Stelling},
\newblock \bibinfo{title}{A subgrid based approach for morphodynamic
  modelling},
\newblock \bibinfo{journal}{Adv. Water Resour.} \bibinfo{volume}{93}
  (\bibinfo{year}{2016}) \bibinfo{pages}{105--117}.
\bibitem[{Danilov(2013)}]{danilov13}
\bibinfo{author}{S.~Danilov},
\newblock \bibinfo{title}{Ocean modeling on unstructured meshes},
\newblock \bibinfo{journal}{Ocean Modell.} \bibinfo{volume}{69}
  (\bibinfo{year}{2013}) \bibinfo{pages}{195–210}.
\bibitem[{Danilov et~al.(2013)Danilov, Wang, Sidorenko, Timmermann, Wekerle,
  Haid, and Wang}]{danilov13c}
\bibinfo{author}{S.~Danilov}, \bibinfo{author}{Q.~Wang},
  \bibinfo{author}{D.~Sidorenko}, \bibinfo{author}{R.~Timmermann},
  \bibinfo{author}{C.~Wekerle}, \bibinfo{author}{V.~Haid},
  \bibinfo{author}{X.~Wang},
\newblock \bibinfo{title}{Multiresolution modeling of large-scale ocean
  circulation},
\newblock \bibinfo{organization}{ECMWF Seminar on Num. Meth. for Atmos. and
  Ocean Modell.}, \bibinfo{year}{2013}.
\bibitem[{Casulli and Zanolli(2002)}]{casulli02}
\bibinfo{author}{V.~Casulli}, \bibinfo{author}{P.~Zanolli},
\newblock \bibinfo{title}{Semi-implicit numerical modeling of nonhydrostatic
  free-surface flows for environmental problems},
\newblock \bibinfo{journal}{Math. Comp. Modell.} \bibinfo{volume}{36}
  (\bibinfo{year}{2002}) \bibinfo{pages}{1131--1149}.
\bibitem[{Wang et~al.(2009)Wang, Fringer, Giddings, and Fong}]{wang09}
\bibinfo{author}{B.~Wang}, \bibinfo{author}{O.~Fringer},
  \bibinfo{author}{S.~Giddings}, \bibinfo{author}{D.~Fong},
\newblock \bibinfo{title}{High-resolution simulations of a macrotidal estuary
  using {SUNTANS}},
\newblock \bibinfo{journal}{Ocean Modell.} \bibinfo{volume}{26}
  (\bibinfo{year}{2009}) \bibinfo{pages}{60--85}.
\bibitem[{Cui et~al.(2010)Cui, Pietrzak, and Stelling}]{cui10}
\bibinfo{author}{H.~Cui}, \bibinfo{author}{J.~Pietrzak},
  \bibinfo{author}{G.~Stelling},
\newblock \bibinfo{title}{A {F}{V} analogue of the ${P}_1^{NC}-{P}_1$ {F}{E}:
  with accurate flooding and drying},
\newblock \bibinfo{journal}{Ocean Modell.} \bibinfo{volume}{35}
  (\bibinfo{year}{2010}) \bibinfo{pages}{16--30}.
\bibitem[{Cui et~al.(2012)Cui, Pietrzak, and Stelling}]{cui12}
\bibinfo{author}{H.~Cui}, \bibinfo{author}{J.~Pietrzak},
  \bibinfo{author}{G.~Stelling},
\newblock \bibinfo{title}{Improved efficiency of a non-hydrostatic,
  unstructured grid, {F}{V} model},
\newblock \bibinfo{journal}{Ocean Modell.} \bibinfo{volume}{54-55}
  (\bibinfo{year}{2012}) \bibinfo{pages}{55--67}.
\bibitem[{Greenberg et~al.(2005)Greenberg, Shore, Page, and Dowd}]{greenberg05}
\bibinfo{author}{D.~A. Greenberg}, \bibinfo{author}{J.~A. Shore},
  \bibinfo{author}{F.~H. Page}, \bibinfo{author}{M.~Dowd},
\newblock \bibinfo{title}{A {F}{E} circulation model for embayments with drying
  intertidal areas and its application to the {Q}uoddy region of the {B}ay of
  {F}undy},
\newblock \bibinfo{journal}{Ocean Modell.} \bibinfo{volume}{10}
  (\bibinfo{year}{2005}) \bibinfo{pages}{211--231}.
\bibitem[{Dietrich et~al.(2006)Dietrich, Kolar, and Westerink}]{dietrich06}
\bibinfo{author}{J.~C. Dietrich}, \bibinfo{author}{R.~L. Kolar},
  \bibinfo{author}{J.~J. Westerink},
\newblock \bibinfo{title}{Refinements in continuous {G}alerkin {W}{D}
  algorithms},
\newblock in: \bibinfo{booktitle}{Estuarine and Coastal Modeling (2005)},
  \bibinfo{publisher}{American Soc. of Civil Eng. ({ASCE})},
  \bibinfo{year}{2006}.
\bibitem[{Bates and Hervouet(1999)}]{bates99}
\bibinfo{author}{P.~D. Bates}, \bibinfo{author}{J.-M. Hervouet},
\newblock \bibinfo{title}{A new method for moving-boundary hydrodynamic
  problems in shallow water},
\newblock \bibinfo{journal}{Proc. R. Soc. Lond. A} \bibinfo{volume}{455}
  (\bibinfo{year}{1999}) \bibinfo{pages}{3107--3128}.
\bibitem[{Funke et~al.(2011)Funke, Pain, Kramer, and Piggott}]{funke11}
\bibinfo{author}{S.~Funke}, \bibinfo{author}{C.~Pain},
  \bibinfo{author}{S.~Kramer}, \bibinfo{author}{M.~Piggott},
\newblock \bibinfo{title}{A {W}{D} algorithm with a combined
  pressure/free-surface formulation for non-hydrostatic models},
\newblock \bibinfo{journal}{Adv. Water Resour.} \bibinfo{volume}{34}
  (\bibinfo{year}{2011}) \bibinfo{pages}{1483--1495}.
\bibitem[{Piggott et~al.(2008)Piggott, Gorman, Pain, Allison, Candy, Martin,
  and Wells}]{piggott08}
\bibinfo{author}{M.~D. Piggott}, \bibinfo{author}{G.~J. Gorman},
  \bibinfo{author}{C.~C. Pain}, \bibinfo{author}{P.~A. Allison},
  \bibinfo{author}{A.~S. Candy}, \bibinfo{author}{B.~T. Martin},
  \bibinfo{author}{M.~R. Wells},
\newblock \bibinfo{title}{A new computational framework for multi-scale ocean
  modelling based on adapting unstructured meshes},
\newblock \bibinfo{journal}{Int. J. Numer. Meth. Fl.} \bibinfo{volume}{56}
  (\bibinfo{year}{2008}) \bibinfo{pages}{1003--1015}.
\bibitem[{Labeur and Pietrzak(2005)}]{labeur05}
\bibinfo{author}{R.~J. Labeur}, \bibinfo{author}{J.~D. Pietrzak},
\newblock \bibinfo{title}{A fully 3{D} unstructured grid non-hydrostatic {F}{E}
  coastal model},
\newblock \bibinfo{journal}{Ocean Modell.} \bibinfo{volume}{10}
  (\bibinfo{year}{2005}) \bibinfo{pages}{51--67}.
\bibitem[{Chorin(1967)}]{chorin67}
\bibinfo{author}{A.~Chorin},
\newblock \bibinfo{title}{A numerical method for solving incompressible viscous
  flow problems},
\newblock \bibinfo{journal}{J. Comput. Phys.} \bibinfo{volume}{2}
  (\bibinfo{year}{1967}) \bibinfo{pages}{12--26}.
\bibitem[{Gresho et~al.(1984)Gresho, Chan, Lee, and Upson}]{gresho84}
\bibinfo{author}{P.~M. Gresho}, \bibinfo{author}{S.~T. Chan},
  \bibinfo{author}{R.~L. Lee}, \bibinfo{author}{C.~D. Upson},
\newblock \bibinfo{title}{A modified {F}{E} method for solving the
  time-dependent, incompressible {N}avier-{S}tokes equations. {P}art 1:
  {T}heory},
\newblock \bibinfo{journal}{Int. J. Numer. Meth. Fl.} \bibinfo{volume}{4}
  (\bibinfo{year}{1984}) \bibinfo{pages}{557--598}.
\bibitem[{Iserles(2012)}]{iserles12}
\bibinfo{author}{A.~Iserles}, \bibinfo{title}{A First Course in the Numerical
  Analysis of Differential Equations}, Cambridge texts in Applied Mathematics,
  \bibinfo{publisher}{Cambridge University Press}, \bibinfo{year}{2012}.
\bibitem[{Courant et~al.(1928)Courant, Friedrichs, and Lewy}]{courant28}
\bibinfo{author}{R.~Courant}, \bibinfo{author}{K.~Friedrichs},
  \bibinfo{author}{H.~Lewy},
\newblock \bibinfo{title}{{\"U}ber die partiellen {D}ifferenzengleichungen der
  mathematischen {P}hysik},
\newblock \bibinfo{journal}{Mathematische Annalen} \bibinfo{volume}{100}
  (\bibinfo{year}{1928}) \bibinfo{pages}{32--74}.
\bibitem[{Kramer et~al.(2010)Kramer, Cotter, and Pain}]{kramer10}
\bibinfo{author}{S.~Kramer}, \bibinfo{author}{C.~Cotter},
  \bibinfo{author}{C.~Pain},
\newblock \bibinfo{title}{Solving the poisson equation on small aspect ratio
  domains using unstructured meshes},
\newblock \bibinfo{journal}{Ocean Modell.} \bibinfo{volume}{35}
  (\bibinfo{year}{2010}) \bibinfo{pages}{253--263}.
\bibitem[{Shchepetkin and McWilliams(2005)}]{shchepetkin05}
\bibinfo{author}{A.~F. Shchepetkin}, \bibinfo{author}{J.~C. McWilliams},
\newblock \bibinfo{title}{The regional oceanic modeling system ({R}{O}{M}{S}):
  {A} split-explicit, free-surface, topography-following-coordinate oceanic
  model},
\newblock \bibinfo{journal}{Ocean Modell.} \bibinfo{volume}{9}
  (\bibinfo{year}{2005}) \bibinfo{pages}{347--404}.
\bibitem[{Maddison et~al.(2011)Maddison, Marshall, Pain, and
  Piggott}]{maddison11}
\bibinfo{author}{J.~Maddison}, \bibinfo{author}{D.~Marshall},
  \bibinfo{author}{C.~Pain}, \bibinfo{author}{M.~Piggott},
\newblock \bibinfo{title}{Accurate representation of geostrophic and
  hydrostatic balance in unstructured mesh {F}{E} ocean modelling},
\newblock \bibinfo{journal}{Ocean Modell.} \bibinfo{volume}{39}
  (\bibinfo{year}{2011}) \bibinfo{pages}{248--261}.
\bibitem[{Karniadakis and Sherwin(1999)}]{karniadakis05}
\bibinfo{author}{G.~Karniadakis}, \bibinfo{author}{S.~J. Sherwin},
  \bibinfo{title}{Spectral/hp Element Methods for CFD},
  \bibinfo{publisher}{Oxford University Press}, \bibinfo{year}{1999}.
\bibitem[{Hughes et~al.(1998)Hughes, Feij{\'{o}}o, Mazzei, and
  Quincy}]{hughes98}
\bibinfo{author}{T.~J. Hughes}, \bibinfo{author}{G.~R. Feij{\'{o}}o},
  \bibinfo{author}{L.~Mazzei}, \bibinfo{author}{J.-B. Quincy},
\newblock \bibinfo{title}{The variational multiscale method{\textemdash}a
  paradigm for computational mechanics},
\newblock \bibinfo{journal}{Comput. Methods Appl. Mech.
  Eng.} \bibinfo{volume}{166} (\bibinfo{year}{1998})
  \bibinfo{pages}{3--24}.
\bibitem[{Candy(2008)}]{candy08}
\bibinfo{author}{A.~S. Candy}, \bibinfo{title}{Subgrid scale modelling of
  transport processes}, Ph.D. thesis, Imperial College London,
  \bibinfo{year}{2008}.
\bibitem[{Candy(2017)}]{candybrep}
\bibinfo{author}{A.~S. Candy},
\newblock \bibinfo{title}{A consistent approach to unstructured mesh generation
  for geophysical models},
\newblock \bibinfo{journal}{In review.}  (\bibinfo{year}{2017}).
\bibitem[{Candy et~al.(2014)Candy, Avdis, Hill, Gorman, and Piggott}]{candy14}
\bibinfo{author}{A.~S. Candy}, \bibinfo{author}{A.~Avdis},
  \bibinfo{author}{J.~Hill}, \bibinfo{author}{G.~J. Gorman},
  \bibinfo{author}{M.~D. Piggott},
\newblock \bibinfo{title}{Integration of geographic information system
  frameworks into domain discretisation and meshing processes for geophysical
  models},
\newblock \bibinfo{journal}{Geosci. Model Dev. Discuss.} \bibinfo{volume}{7}
  (\bibinfo{year}{2014}) \bibinfo{pages}{5993--6060}.
\bibitem[{Kleptsova et~al.(2010)Kleptsova, Stelling, and
  Pietrzak}]{kleptsova2010}
\bibinfo{author}{O.~Kleptsova}, \bibinfo{author}{G.~S. Stelling},
  \bibinfo{author}{J.~D. Pietrzak},
\newblock \bibinfo{title}{An accurate momentum advection scheme for a z-level
  coordinate models},
\newblock \bibinfo{journal}{Ocean Dyn.} \bibinfo{volume}{60}
  (\bibinfo{year}{2010}) \bibinfo{pages}{1447--1461}.
\bibitem[{Bleck(2002)}]{bleck2002}
\bibinfo{author}{R.~Bleck},
\newblock \bibinfo{title}{An oceanic general circulation model framed in hybrid
  isopycnic-cartesian coordinates},
\newblock \bibinfo{journal}{Ocean Modell.} \bibinfo{volume}{4}
  (\bibinfo{year}{2002}) \bibinfo{pages}{55--88}.
\bibitem[{Burchard and Petersen(1997)}]{burchard1997}
\bibinfo{author}{H.~Burchard}, \bibinfo{author}{O.~Petersen},
\newblock \bibinfo{title}{Hybridization between $\sigma$- and z- co-ordinates
  for improving the internal pressure gradient calculation in marine models
  with steep bottom slopes},
\newblock \bibinfo{journal}{Int. J. Numer. Meth. Fl.} \bibinfo{volume}{25}
  (\bibinfo{year}{1997}) \bibinfo{pages}{1003--1023}.
\bibitem[{Farrell et~al.(2011)Farrell, Piggott, Gorman, Ham, Wilson, and
  Bond}]{farrell11}
\bibinfo{author}{P.~E. Farrell}, \bibinfo{author}{M.~D. Piggott},
  \bibinfo{author}{G.~J. Gorman}, \bibinfo{author}{D.~A. Ham},
  \bibinfo{author}{C.~R. Wilson}, \bibinfo{author}{T.~M. Bond},
\newblock \bibinfo{title}{Automated continuous verification for numerical
  simulation},
\newblock \bibinfo{journal}{Geosci. Model Dev.} \bibinfo{volume}{4}
  (\bibinfo{year}{2011}) \bibinfo{pages}{435--449}.
\bibitem[{Cotter et~al.(2009)Cotter, Ham, and Pain}]{cotter09}
\bibinfo{author}{C.~J. Cotter}, \bibinfo{author}{D.~A. Ham},
  \bibinfo{author}{C.~C. Pain},
\newblock \bibinfo{title}{A mixed discontinuous/continuous {F}{E} pair for
  shallow-water ocean modelling},
\newblock \bibinfo{journal}{Ocean Modell.} \bibinfo{volume}{26}
  (\bibinfo{year}{2009}) \bibinfo{pages}{86--90}.
\bibitem[{Hestenes and Stiefel(1952)}]{hestenes52}
\bibinfo{author}{M.~R. Hestenes}, \bibinfo{author}{E.~Stiefel},
\newblock \bibinfo{title}{Methods of conjugate gradients for solving linear
  systems} \bibinfo{volume}{49} (\bibinfo{year}{1952})
  \bibinfo{pages}{409--436}.
\bibitem[{Young(1971)}]{young71}
\bibinfo{author}{D.~M. Young}, \bibinfo{title}{{I}terative {S}olution of
  {L}arge {L}inear {S}ystems}, \bibinfo{publisher}{Academic Press},
  \bibinfo{year}{1971}.
\bibitem[{Saad and Schultz(1986)}]{saad86}
\bibinfo{author}{Y.~Saad}, \bibinfo{author}{M.~Schultz},
\newblock \bibinfo{title}{{GMRES}: a generalized minimal residual algorithm for
  solving nonsymmetric linear systems},
\newblock \bibinfo{journal}{SIAM J. Sci. Stat. Comput.} \bibinfo{volume}{7}
  (\bibinfo{year}{1986}) \bibinfo{pages}{856--869}.
\bibitem[{Balay et~al.(1997)Balay, Gropp, McInnes, and Smith}]{petsc}
\bibinfo{author}{S.~Balay}, \bibinfo{author}{W.~D. Gropp},
  \bibinfo{author}{L.~C. McInnes}, \bibinfo{author}{B.~F. Smith},
\newblock \bibinfo{title}{Efficient management of parallelism in object
  oriented numerical software libraries},
\newblock in: \bibinfo{editor}{E.~Arge}, \bibinfo{editor}{A.~M. Bruaset},
  \bibinfo{editor}{H.~P. Langtangen} (Eds.), \bibinfo{booktitle}{Modern
  Software Tools in Scientific Computing}, \bibinfo{publisher}{Birkh{\"{a}}user
  Press}, \bibinfo{year}{1997}, pp. \bibinfo{pages}{163--202}.
\bibitem[{Gourgue et~al.(2009)Gourgue, Comblen, Lambrechts, K{\"a}rn{\"a},
  Legat, and Deleersnijder}]{gourgue09}
\bibinfo{author}{O.~Gourgue}, \bibinfo{author}{R.~Comblen},
  \bibinfo{author}{J.~Lambrechts}, \bibinfo{author}{T.~K{\"a}rn{\"a}},
  \bibinfo{author}{V.~Legat}, \bibinfo{author}{E.~Deleersnijder},
\newblock \bibinfo{title}{A flux-limiting {W}{D} method for {F}{E}
  shallow-water models, with application to the {S}cheldt {E}stuary},
\newblock \bibinfo{journal}{Adv. Water Resour.} \bibinfo{volume}{32}
  (\bibinfo{year}{2009}) \bibinfo{pages}{1726 -- 1739}.
\bibitem[{Thacker(1981)}]{thacker81}
\bibinfo{author}{W.~C. Thacker},
\newblock \bibinfo{title}{Some exact solutions to the nonlinear shallow-water
  wave equations},
\newblock \bibinfo{journal}{J. Fluid. Mech.} \bibinfo{volume}{107}
  (\bibinfo{year}{1981}) \bibinfo{pages}{499}.
\bibitem[{Casulli and Zanolli(2007)}]{casulli07}
\bibinfo{author}{V.~Casulli}, \bibinfo{author}{P.~Zanolli},
\newblock \bibinfo{title}{Comparing analytical and numerical solution of
  nonlinear 2 and 3{D} hydrostatic flows},
\newblock \bibinfo{journal}{Int. J. Numer. Meth. Fl.} \bibinfo{volume}{53}
  (\bibinfo{year}{2007}) \bibinfo{pages}{1049--1062}.
\bibitem[{Acheson(1990)}]{acheson90}
\bibinfo{author}{D.~Acheson}, \bibinfo{title}{Elementary fluid dynamics},
  Oxford Appl. Math. And Comp. Sci. Series, \bibinfo{publisher}{Oxford
  University Press}, \bibinfo{year}{1990}.
\bibitem[{Brenner and Scott(1994)}]{brenner94}
\bibinfo{author}{S.~Brenner}, \bibinfo{author}{R.~Scott}, \bibinfo{title}{The
  Mathematical Theory of Finite Element Methods},
  \bibinfo{publisher}{Springer-Verlag§}, \bibinfo{year}{1994}.

\end{thebibliography}
\endgroup
\end{document}